\documentclass[12pt]{article}
 \usepackage[top=1in, bottom=1in, left=0.8in, right=0.8in]{geometry}
\usepackage{graphicx}
\usepackage{appendix}
\usepackage{authblk}
\usepackage[numbers,sort&compress]{natbib}
\usepackage{natbib}
 \usepackage{pdflscape}
\usepackage{amssymb}\usepackage{amsmath,amsthm}
\usepackage{amsmath} \usepackage{multirow}

\renewcommand{\theequation}{\thesection.\arabic{equation}}

\numberwithin{equation}{section}
\usepackage{amscd,epstopdf}
\usepackage{graphicx}\usepackage{multirow}
\usepackage{float}
\usepackage{subcaption}
\usepackage{afterpage} \usepackage{rotating}
\usepackage{enumerate}

 \usepackage{setspace}
\usepackage{indentfirst}
\usepackage{graphicx}
\usepackage{bibentry}

\begin{document}

\title{Expectation-Maximizing Network Reconstruction and  Most Applicable Network Types Based on Binary Time Series Data}

\author[1]{Kaiwei Liu}
\author[1,\thanks{Corresponding author with E-mail address:
XLV@bjtu.edu.cn}]{Xing L\"{u}}
\author[2]{Fei Gao}
\author[2,3]{Jiang Zhang}

\affil[1]{{\footnotesize Department of Mathematics, Beijing Jiaotong
University, Beijing 100044, China}} \affil[2]{{\footnotesize School
of Systems Science, Beijing Normal University, Beijing 100875,
China}} \affil[3]{{\footnotesize Swarma Research, Beijing 102399, China}}

 \renewcommand\Authands{ and }
 \date{}
 \maketitle

\begin{abstract}
Based on the binary time series data of social infection dynamics, we propose a general framework to reconstruct the 2-simplex complexes with two-body and three-body interactions by combining the maximum likelihood estimation in statistical inference and introducing the expectation maximization. In order to improve the code running efficiency, the whole algorithm adopts vectorization expressions. Through the inference of maximum likelihood estimation, the vectorization expression of the edge existence probability can be obtained, and through the probability matrix, the adjacency matrix of the network can be estimated. We apply a two-step scheme to improve the effectiveness of network reconstruction while reducing the amount of computation significantly. The framework has been tested on different types of complex networks. Among them, four kinds of networks achieve high reconstruction effectiveness. Besides, we study the influence of noise data or random interference and prove the robustness of the framework, then the effects of two kinds of hyper-parameters on the experimental results are tested. Finally, we analyze which type of network is more suitable for this framework, and propose methods to improve the effectiveness of the experimental results.
\end{abstract}


\section{Introduction}
In reality, many systems can be represented by models of complex networks \cite{Baggio2010}. The information propagation in the system can be simulated by the propagation dynamics model \cite{Utterback1975}. However, the inherent network system in nature cannot be directly observed by humans, and we can only deduce the underlying network structure based on phenomena. The most intuitive form of expression for observed phenomena is data. Once we have the data, we can use the knowledge and methods of statistics, network science and engineering to reconstruct the network \cite{Wang2016}. Some basic concepts of these three fields are described in detail below.

\subsection{Maximum likelihood estimation and expectation maximization estimation}
An important part of statistics is to infer the statistical information and potential characteristics of data sets, including parameter estimation. Maximum likelihood estimation (MLE) \cite{Dempster1977, Meng1991, Qian1992, Hermanns2020} is one of the most commonly used methods in parameter estimation. It is often used as a reference value in real data analysis. In parameter estimation, we often optimize the effectiveness of estimation by adopting the algorithm for finding the expectation on the basis of parameter estimation, such as the expectation-maximization (EM) algorithm. Combining Markov algorithm makes the operation more reasonable and convenient. Similar optimization methods can be used in other types of parameter estimation, such as E-Bayesian estimation based on Bayesian estimation \cite{Han2020, Han2019b, Liu2021, Zhang2021}. Reconstructing the network by using E-Bayesian estimation can also be attempted in the future. The essence of parameter estimation is to estimate the probability of different events. Applying the parameter estimation methods in statistics to complex networks is the latest research direction.

\subsection{Reconstruction of complex networks}
\begin{figure}[htbp]
    \begin{minipage}[c]{0.5\textwidth}
        \centering
        \includegraphics[width=1\textwidth,trim=50 40 50 45,clip]{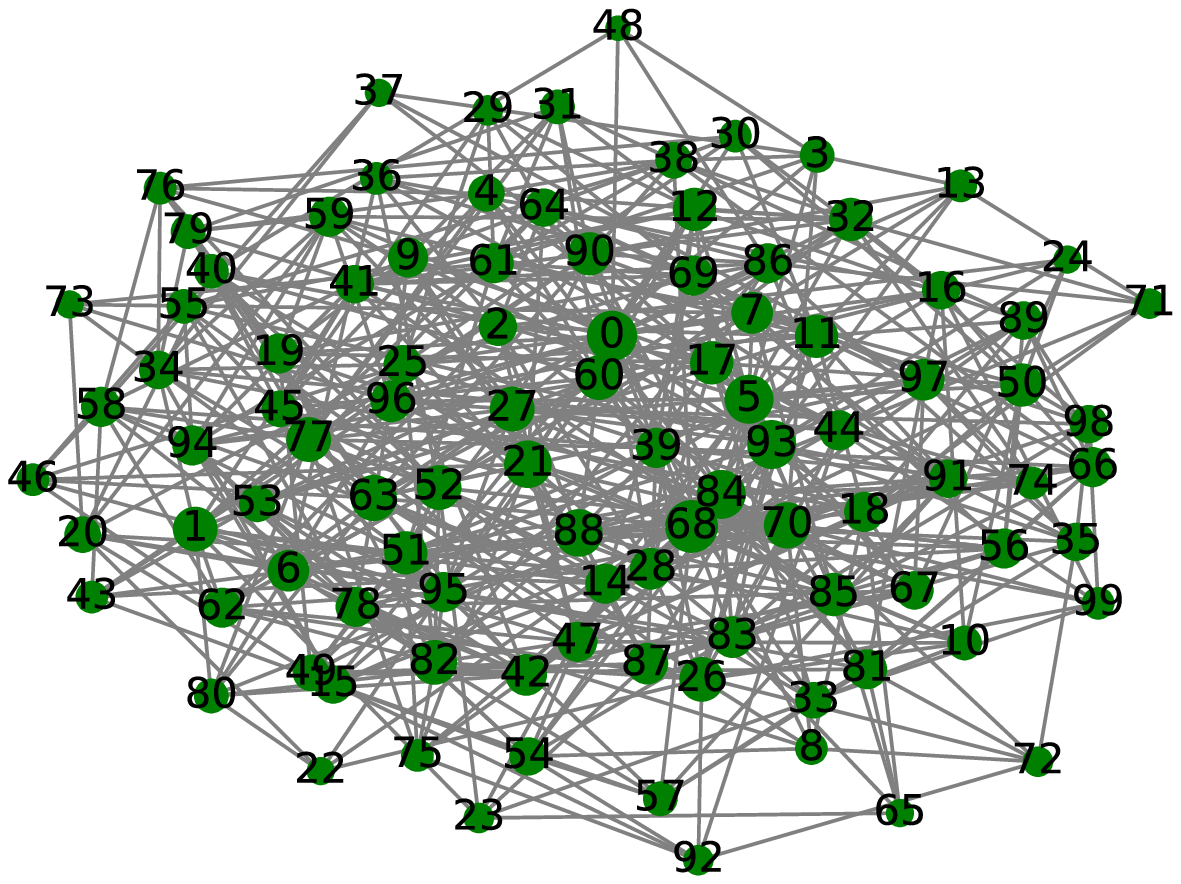}
        \centerline{(a)}
    \end{minipage}
    \begin{minipage}[c]{0.5\textwidth}
        \centering
        \includegraphics[width=1\textwidth,trim=50 40 50 45,clip]{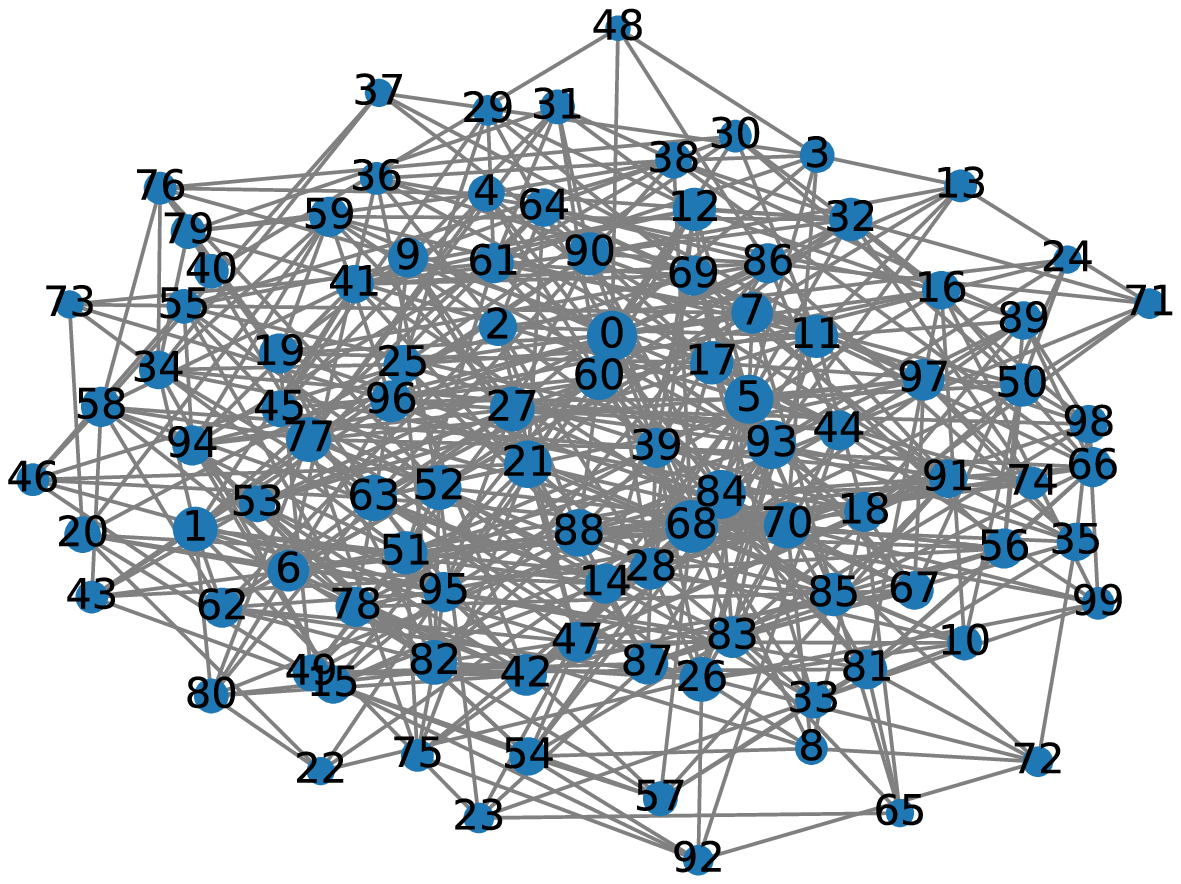}
        \centerline{(b)}
\end{minipage}
    \caption{Example of a binomial distribution network, (a) the original network and (b) the reconstructed network. The numbers in the figure represent the serial numbers of the network nodes, and the serial numbers of the network nodes correspond to the serial numbers of the columns or rows representing the nodes in the matrix.}
    \label{network_example}
\end{figure}

The purpose of complex network reconstruction is to obtain the recurrence of the original network from the prediction in which we reproduce the community of nodes and regenerate edge or even hyper-edge connections in the network \cite{Wang2016, Newman2016, DeBacco2017, Iacopini2019}. At present, a very common algorithm is to speculate the existence of edge connectivity based on time series data \cite{Ma2015, Ma2018}. For various types of networks \cite{Barabasi1999, Watts1998, Zhang2016}, the applicability of time series through the propagation process is also different. On the basis of common network reconstruction \cite{Shen2014}, we can introduce high-order networks \cite{Bai2021,Huang2021}, and use the same theoretical method \cite{Wang2016, Newman2016, DeBacco2017, Iacopini2019, Ma2015, Ma2018, Wang2022} to judge the existence of hyper-edges. The effectiveness of the reconstructed network can be adopted to a unified standard \cite{Powers2020}, which we will specify later. In the past two years, people have become more and more interested in hyper-graphs \cite{Iacopini2019, Carletti2020, Carletti2021, FerrazdeArruda2021, Arruda2020, Landry2020, Matamalas2020, Wang2020, Young2021, Wang2022}. Complex networks are widely used as potential and possibly multi-level social structures as they can well describe the connectivity in systems of various properties  \cite{Li2015, Zhang2020, Wu2020}. Figure \ref{network_example} shows an example of the original network and the reconstructed network, and our analysis of the algorithm is based on this case.

\subsection{Simplex complexes}
Although pairwise or node-to-node interaction is a common type in the network, people have realized that high-order interaction is also common and has important influences on graph network structure. A group of nodes in the network may interact at the same time. It is no longer enough to describe the network with the traditional pairwise interaction \cite{Watts1998}. Therefore, higher-order interactions other than pairwise relationships must be considered. Mathematically, higher-order interactions can be described as hyper-graphs or simplex complexes \cite{Shen2014a, Burgio2021, Battiston2020}, those are networks containing higher-order simplex. Hyper-graph is an expression structure that can flexibly simulate the high-order correlation between entities. It has attracted extensive attention in various research fields in recent years. Though graph neural network (GNN) has achieved success in graph representation learning, it is still a challenging task to apply powerful GNN variants to hyper-graphs directly \cite{Bai2021, Huang2021}. Hyper-graph can be represented by the set of hyper-edges, while we mainly use simplex to build high-order interactive networks this time.

The $K$-simplex describes the simultaneous interaction between $k+1$ nodes. Any higher-order simplex can be disassembled into several lower-order simplex. According to the definition, the $k$-order simplex can be disassembled into $(k-1)$-simplex whose quantity is $k+1$. The interaction on a $k$-simplex can be defined as the effect of the $(k-1)$-simplex on a single node. The network composed of simplex is called simplex complex. An article on simplex for social models \cite{Iacopini2019} details the definition of simplex and its relationship to hypergraphs. In addition, related concepts \cite{Zhao2021, Battiston2020} are also widely studied and applied in physics. For specific simplex structure and dynamics, please refer to Appendix \ref{simple}.

\subsection{Dynamic mechanism and simulation method of SIS propagation}

We divide the infection state of nodes on a simple complex into two categories, say as infectious state and susceptible state. Infectious status indicates that the node has infection source or information that can be transmitted. Susceptible status indicates that the node is likely to be infected in the future.  We use $\psi_i^t$ to represent the infected state of node $i$ at time $t$. If node $i$ is infected at time $t$, we specify $\psi_i^t=1$; Similarly, when node $i$ is susceptible, we have  $\psi_i^t=0$.

In the simulation process of this paper, we assume that the probability of all node to node propagation, that is, the probability of interaction on 1-simplex $(i,j)$, is $\beta_ 1$. Similarly, the probability of interaction on 2-simplex $(i,j,k)$ is $\beta_2$. Normally, we specify $\beta_2> \beta_1$ to express that the probability of three-body propagation is higher than that of two-body propagation. For convenience, we can associate them with the simple complex's two-body two-body and three-body average degrees $k_1$ and $k_2$. We specify that $\beta_1= \alpha/k_1$. At the same time $\beta_ 2= \omega/k_2$, where $\alpha$ and $\omega$ are two pre-defined parameters, $\alpha<\omega$. The detailed definition of simplex and the dynamic propagation mechanism on it \cite{Iacopini2019, Battiston2020} can be viewed in the literature on propagation dynamics.

In this way, according to the probability formula, we can know that for a susceptible node $i$,  if there are $n_1$ adjacent nodes in the infected state, the probability that it will be infected by the two-body interactions at the next time point on 1-simplex is
\begin{eqnarray} \label{p1}
    p_1=
    \displaystyle\begin{cases}
        \displaystyle\prod_{j(\psi_j^t=1,j\in\mathcal{N}_1(i))}(1-\beta_1), &n_1\ne0\\
        0,&n_1=0
    \end{cases},
\end{eqnarray}
where $\mathcal{N}_1(i)$ represents the set of 0-simplexes adjacent to $i$, which means, the set of adjacent nodes of $i$. Similarly, for a susceptible node $i$, if $n_2$ adjacent 1-simplexes are in the infected state and can form a 2-simplex with node $i$, the probability of infection by the three-body interaction at the next time point is
\begin{eqnarray} \label{p2}
    p_2=
    \displaystyle\begin{cases}
        \displaystyle\prod_{jk(\psi_{jk}^t=1,jk\in\mathcal{N}_2(i))}(1-\beta_2), &n_2\ne0\\
        0,&n_2=0
    \end{cases},
\end{eqnarray}
where $\mathcal{N}_2(i)$ represents a set of 1-simplexes adjacent to $i$ that can form a 2-simplex with this node, $\psi_{jk}^t=\psi_{k}^t\psi_{j}^t$. Then, for a susceptible node $i$, the probability of being infected at the next time point is
\begin{eqnarray}\label{pp1p2}
    p=1-(1-p_1)(1-p_2).
\end{eqnarray}
We set the proportion of initial infected nodes as $\rho_0$, the probability of the infected node returning to the susceptible state is $\mu$.

In this way, with formulas (\ref{p1})-(\ref{pp1p2}), we can design a complete simulation process to simulate the propagation process on simplex, and then study the mechanism of complex network reconstruction. The reconstruction process based on the maximum likelihood estimation is introduced in the next section.\\

In this paper, complex networks are expressed in the form of simplex complexes. The purpose of this paper is to derive the adjacency matrix and obtain the reconstruction model of the complex network by estimating the probability of edge connection between different nodes with using the maximum likelihood estimation (MLE) and expectation maximization (EM) methods \cite{Dempster1977, Meng1991, Qian1992, Hermanns2020}, which is based on the binary time-series data \cite{Kedem1980} generated by the susceptible-infectious-susceptible (SIS) information propagation model \cite{Wang2019}. On the basis of complex networks, we introduce high-order networks and use the form of the simplex complexes to infer the information transmission between communities \cite{Wang2016, Newman2016, DeBacco2017, Iacopini2019}. The generated probability is compared with a threshold to determine the existence of the edge between two points. Then, we use $F1$ criterion \cite{Powers2020}, which is a judging criterion based on the joint calculation of the existence of real network connections and the prediction of reconstructed network connections, to evaluate the effectiveness (or the accuracy) of the reconstruction.

The innovations of this paper are as follows:
\begin{enumerate}[(1)]
    \item The vectorization expression of the final probability estimation is introduced to increase the code operation efficiency, and the probability matrix generated can be directly transformed into the adjacency matrix of the network.
    \item Different types of networks are generated in the form of the simplex complexes to analyze the changes in different types of networks' reconstruction effectiveness under the same hyper-parameters.
    \item Through the comparison of different types of networks, the relationship between degree variance and reconstruction effectiveness is introduced, and find that the small-world network is the most suitable network for the reconstruction method. The actual social, ecological, and other networks are small-world networks \cite{Watts1998}, so this method is feasible in reality.
    \item At last, we summarize the method to obtain and process network-related data, which can improve the effectiveness of reconfiguration from the direction of network structure. We put forward three optimization methods: decentralization, extending the scale, and optimizate the algorithm.
\end{enumerate}

The outline of this paper is as follows. In Section 2, we will study the operation method of specific probability estimation, matrix expression, threshold calculation, and the selection of evaluation criteria. In Section 3, we conduct specific digital experiments on several different types of networks, take four groups of the best experimental results as demonstration examples, and visualize the generated data. Section 4 analyzes the reasons why network reconstruction performs best in the small-world network in Section 3. Finally, the conclusion is summarized in Section 5.

\section{Reconstruction of simple complex based on MLE}

The complex network reconstruction is to infer whether there are edge connections between nodes in the network from the time series of node infection states. In the reconstruction of ordinary networks \cite{Ma2018} and simplicial complexes  \cite{Wang2022}, the idea of inference has been systematically inferred based on the infection state $\psi_i^t$ of different nodes and times as well as the propagation probability $P_{j\to i}$ between nodes. Although this method is feasible, it will lead to a significant increase in the volumn of calculation when estimating the propagation probability between nodes, especially in the situation the network is complex and the number of nodes is large. For example, if the network with $n$ nodes is calculated separately, there will be $n(n-1)/2$ probabilities calculated. If iteration is added, the solution process will be more complex and the expression will be more cumbersome. At present, better solutions are still being explored. In order to have a more concise model, we introduce the concept of vectorization. Eventually, the probabilities and infection states in the experiments are expressed in the form of vectors and matrices. 

\subsection{Derivation of propagation probability and vectorized expression}

\begin{figure}[htbp]
    \begin{minipage}[t]{\linewidth}
        \centering
        \includegraphics[width=0.8\linewidth]{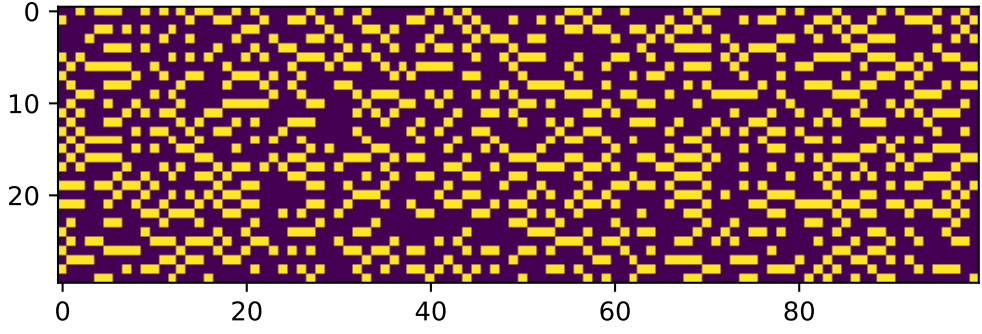}
    \end{minipage}
    \caption{Time series matrix $\Psi_1=(\psi_i^{t_m})_{T\times n}$, yellow means $\psi_j^{t_m}=1$, purple means $\psi_j^{t_m}=0$, a row of the matrix can represent the infection status of all nodes in the network at a certain time point $t_m$, and a column represents the time series of the infection status of a certain node. Due to the excessive number of matrix columns, we only take the first 30 time points for display.}
    \label{psi}
\end{figure}

As defined in Refs.\cite{Iacopini2019, Battiston2020}, 0-simplex can be regarded as a node, and 1-simplex can be regarded as a connected edge. 2-simplex is defined as a three body community in which 1-simplex can transfer information to a single node. We use matrix $\Psi_1=(\psi_i^{t_m})_{T\times n}$ to represent the time series of infection status of the whole network or simplex complex. The number of rows $T$ of the matrix represents the number of iteration steps of the propagation process, and the number of columns $n$ represents the number of nodes. A single column in the matrix can be regarded as the time series of infection status of a single node. The yellow matrix elements in Fig.\ref{psi} represent $\psi_i^{t_m}=1$ and the purple represent $\psi_i^{t_m}=0$, the time series matrix represented by this matrix is generated by using the simulation method in subsection 1.4.

On the basis of matrix $\Psi_1$, we can generate the matrix of infection state of 1-simplex, which is represented by time series matrix $\Psi_2=(\psi_{jk}^{t_m})_{T\times n_1}$. If both nodes $j$ and $k$ are in the infected state, we specify $\psi_{jk}^{t_m}=1$. As long as one of them is in the susceptible state, $\psi_{jk}^{t_m}=0$. Therefore, we can express the relationship between infection status of 1-simplex and 0-simplex, or the relationship between edge and node infection status, as $\psi_{jk}^{t_m}=\psi_{k}^{t_m}\psi_{j}^{t_m}$. In this way, we can better judge whether the 2-simplex containing these three nodes exists by testing whether the propagation of 1-simplex to nodes is generated by the propagation probability $P_{jk\to i}$ of 1-simplex points to a single node. As this paper mainly studies undirected graphs, $(j,k)$ and $(k,j)$ have the same meaning, $\psi_{jk}^{t_m}$ and $\psi_{kj}^{t_m}$ also have the same meaning.

When training our model, we often hope that we can process the whole small batch of samples at the same time. In order to achieve this target, we need to vectorize the calculation so that we can use the linear algebra library instead of writing expensive and inefficient \emph{for} loops in Python. For the elements in the matrix, we can also split the operation, which is more verbose, but expresses the specific elements more clearly \cite{Ma2018, Wang2022}. Refer to Appendix \ref{ML} for the specific calculation process of the maximum likelihood estimation.

In order to meet the required expression, we introduce matrices $Pt_1=(P_j^i)_{n\times n}$ and $Pt_2=(P_{jk}^i)_{n\times n_1}$. In the two matrices, the number of rows $n$ represents that of nodes $i$ that may be infected by neighboring nodes, resulting in the transformation of their own state from susceptible state to infected state. The column numbers $n$ and $n_1$ represent the number of nodes $j$ or node groups $(j,k)$ that are infected and may convert other nodes into infected states. The parameters in the iteration process are expressed by $\rho_{i,j}^{t_m}$, $\rho_{i,jk}^{t_m}$ and $\rho_{\varepsilon_i}^{t_m}$. Here, we use three-dimensional matrices $R_1=(\rho_{i,j}^{t_m})_{n\times n\times T}$, $R_2=(\rho_{i,jk}^{t_m})_{n\times n_1\times T}$ and two-dimensional matrix $R_{\varepsilon}=(\rho_{\varepsilon_i}^{t_m})_{n\times T}$ to represent the matrix of iteration parameters. On each two-dimensional matrix where node $t_m$ of the three-dimensional matrix is located, the number of rows $n$ represents the number of nodes $i$ that may be propagated to cause the transformation from susceptible state to infectious state, the number of columns  $n$ and $n_1$ represent the number of nodes $j$ or node groups $(j,k)$ that are in infectious state and may convert other nodes to infectious state, the third dimension represents the time dimension corresponding to the time series. If there is a connecting edge between $j$ and $k$, $(j,k)$ can be regarded as a 1-simplex.

The following probability estimation is calculated by using vectorization expression. All operations in this subsection are derived from maximum likelihood estimation and EM algorithm. The meaning of each element in the vectors and the derivation method by EM algorithm can be referred to the relevant literature for non-vectorized expressions \cite{Ma2018, Wang2022}. The results of vectorization calculation are consistent with the results of individual calculation for each element. Therefore, this method can be used. For specific matrix decomposition, we can refer to Appendix \ref{vector}. From Appendix \ref{vector} we can see the elements of matrices $P_1$, $P_2$, $Pt_1$, $Pt_2$, $R_1$ and $R_2$ as well as their noise and the median of the first step of the two-step iteration.

\subsubsection{One-step iterative algorithm}

\begin{figure}[htbp]
    \begin{minipage}[c]{0.5\textwidth}
        \centering
        \includegraphics[width=1\textwidth,trim=50 0 60 30,clip]{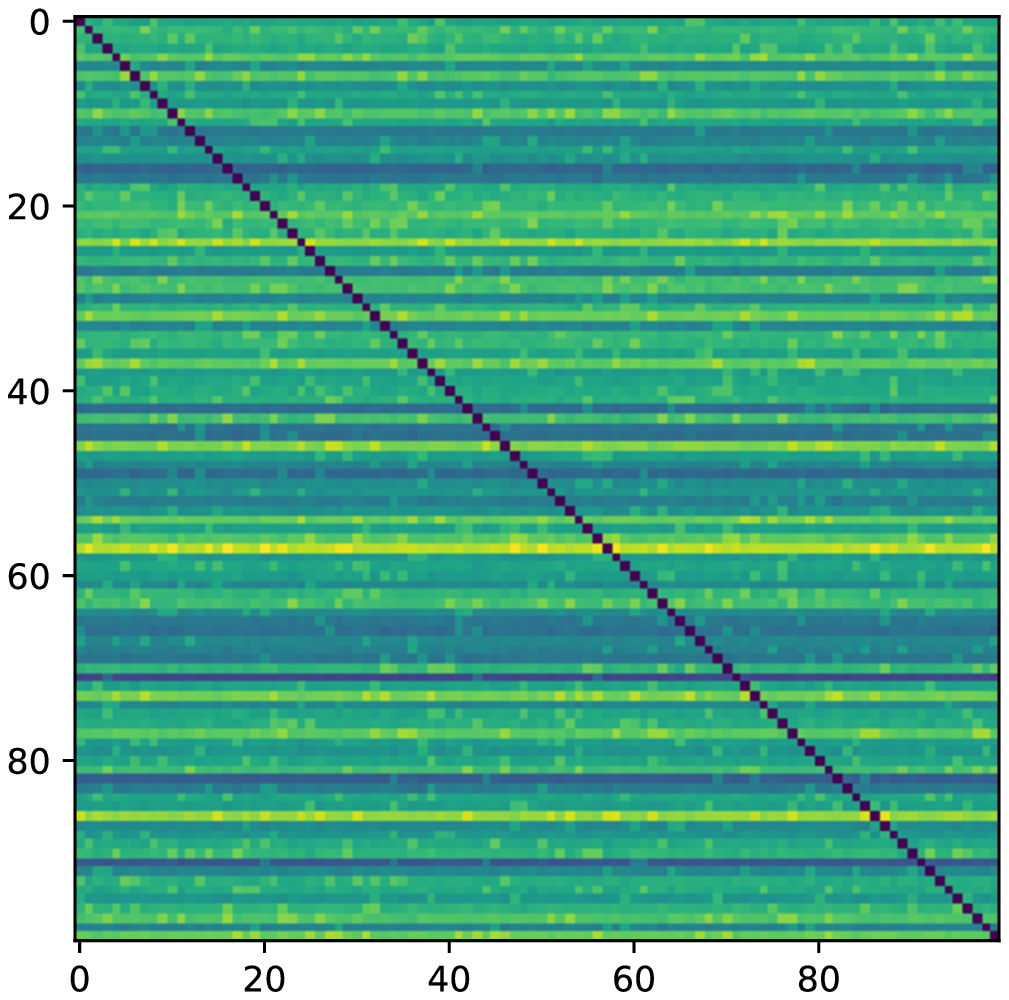}
        \centerline{(a)}
    \end{minipage}
    \begin{minipage}[c]{0.5\textwidth}
        \centering
        \includegraphics[width=1\textwidth,trim=60 0 50 30,clip]{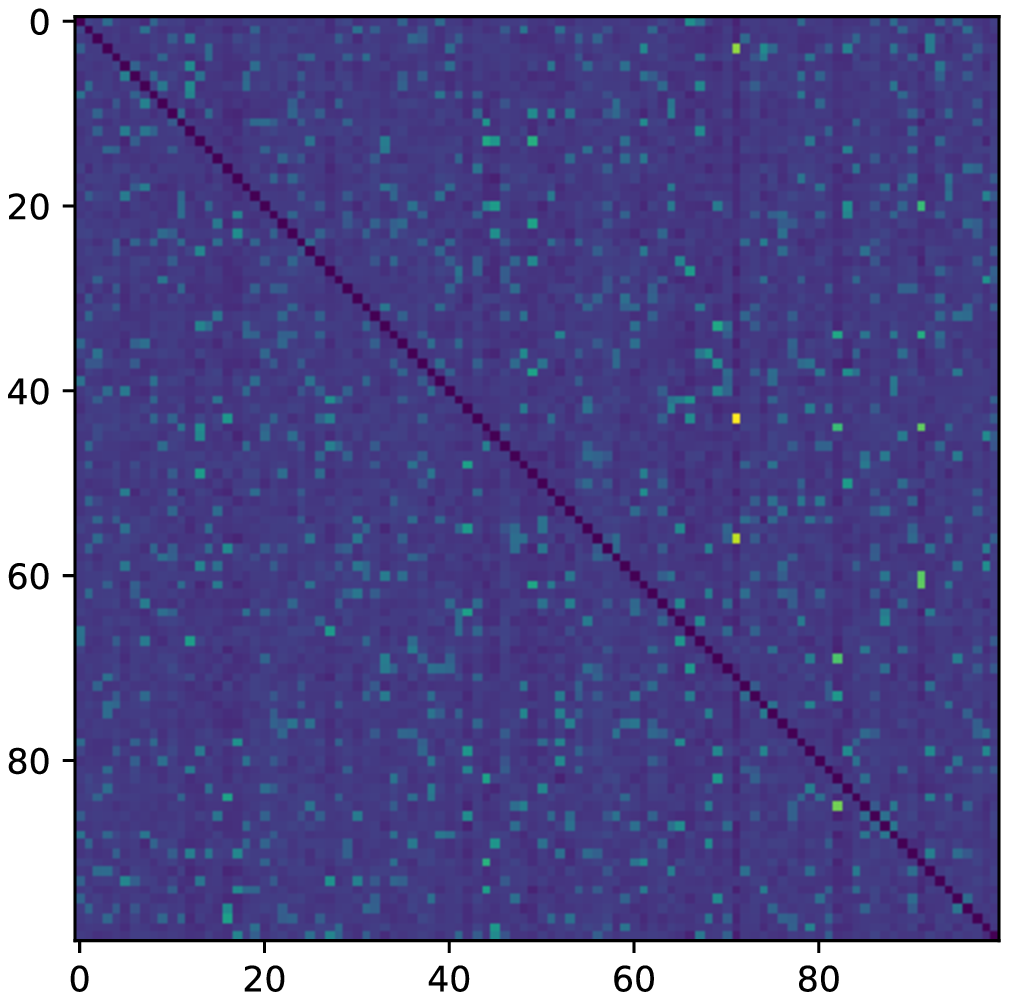}
        \centerline{(b)}
    \end{minipage}
    \caption{Matrix visualization, respectively (a)$Pt_1$, (b)$P_1$. Although the symmetry elements can be vaguely seen in $Pt_1$, they are very inconspicuous. $P_1$ can be approximately regarded as a symmetric matrix, and an adjacency matrix can be generated from it after transformation.}
    \label{P1}
\end{figure}

Iterative indexes and target parameters are very important in the EM algorithm. Using matrices $R_1$, $R_2$ and $R_{\varepsilon}$, we get an intermediate variable
\begin{eqnarray}\label{stm}
    S^{t_m}=(P_1^T \ast Pt_1)\circ\Psi_1^{t_m}+(P_2^T \ast Pt_2)\circ\Psi_2^{t_m},
\end{eqnarray}
where each dimension of $S^{t_m}$ represents the value of $\sum_{j,j\ne i} P_j^i P_{j\to i}\psi_j^{t_m}+\sum_{jk,j,k\ne i}P_{jk}^i P_{jk\to i}\psi_j^{t_m}\psi_k^{t_m}$ for different nodes $i$ and time points $t_m$. $S^{t_m}:n\times 1$, $R_1^{t_m}:n\times n$, $R_2^{t_m}:n\times n_1$, and $R_{\varepsilon}^{t_m}:n\times 1$. So we can get three iterative formulas as
\begin{eqnarray}\label{r1m}
    R_1^{t_m}=[(1_{n\times 1}\circ\Psi_1^{t_m})\ast(P_1^T \ast Pt_1)]/[(Ep+S^{t_m})\circ 1_{1\times n}],
\end{eqnarray}
\begin{eqnarray}\label{r2m}
    R_2^{t_m}=[(1_{n\times 1}\circ\Psi_2^{t_m})\ast(P_2^T \ast Pt_2)]/[(Ep+S^{t_m})\circ 1_{1\times n}],
\end{eqnarray}
\begin{eqnarray}\label{rem}
    R_{\varepsilon}^{t_m}=Ep/[(Ep+S^{t_m})\circ 1_{1\times n}].
\end{eqnarray}
In the matrix calculation, we use $\circ$ to represent matrix multiplication in linear algebra, $\ast$ to represent matrix dot product, and $/$ to represent matrix dot division. Such as matrix $A=(a_{ij})_{m\times  n}$ and $B=(b_{ij})_{m\times n}$, $A\ast B=(a_{ij}b_{ij})_{m\times n}$, $A/ B=(a_{ij}/b_{ij})_{m\times n}$. $P_1=(P_{j\to i})_{n\times n}$, $P_2=(P_{jk\to i})_{n_1\times n}$ are the two probability matrices we want to derive and $Ep=(\varepsilon_i)_n$ is the noise vector. The number of columns $n$ represents the number of nodes $i$ that may be propagated the transition from susceptible state to infected state, and the number of rows $n$ and $n_1$ represent the number of nodes $j$ and node group $(j,k)$ that are infected and may convert other nodes to infected state. If one-step method is adopted, $n_1=n(n-1)/2$. The vectorized probability matrix $P_1$, $P_2$ and noise vector $Ep$ are expressed as follows:
\begin{eqnarray}\label{p1m}
    P_1(:;i)=(S_{11}^i/S_{12}^i)^T,
\end{eqnarray}
\begin{eqnarray}\label{p2m}
    P_2(:;i)=(S_{21}^i/S_{22}^i)^T,
\end{eqnarray}
\begin{eqnarray}\label{pem}
    \varepsilon_i=(S_{\varepsilon_1}^i/S_{\varepsilon_2}^i)^T,
\end{eqnarray}
where $P_1(:;i)$ and $P_2(:;i)$ represents the vectors composed of the elements in column $i$ of the matrix, respectively, and

\begin{eqnarray} \label{p1s}
    \displaystyle\begin{cases}
        \displaystyle S_{11}^i=1_{1\times{(T-1)}}\circ\{R_1^{1:(T-1)}(i;:)\ast[(I_{\psi_1(1:T-1,i)=0}\ast\Psi_1(2:T,i))\circ 1_{1\times n}]\}\\
        S_{12}^i=1_{1\times{(T-1)}}\circ\{\Psi_1 (1:T-1,:)\ast[1_{(T-1)\times1}\circ Pt_1(i,:)]\ast[I_{\Psi_1 (1:T-1,i)=0}\circ 1_{1\times  n}]\}
    \end{cases},
\end{eqnarray}
\begin{eqnarray}\label{p2s}
        \displaystyle\begin{cases}
        \displaystyle S_{21}^i=1_{1\times{(T-1)}}\circ\{R_2^{1:(T-1)}(i;:)\ast[(I_{\psi_1(1:T-1,i)=0}\ast\Psi_1(2:T,i))\circ 1_{1\times n_1}]\}\\
        S_{22}^i=1_{1\times{(T-1)}}\circ\{\Psi_2 (1:T-1,:)\ast[1_{(T-1)\times1}\circ Pt_2(i,:)]\ast[I_{\Psi_1 (1:T-1,i)=0}\circ 1_{1\times  n_1}]\}
    \end{cases},
\end{eqnarray}
\begin{eqnarray} \label{pes}
    \displaystyle\begin{cases}
    \displaystyle S_{\varepsilon_1}^i=1_{1\times{(T-1)}}\circ[R_{\varepsilon}^{1:(T-1)}(i)\ast I_{\psi_1(1:T-1,i)=0}\ast\Psi_1(2:T,i)]\\
    S_{\varepsilon_2}^i=1_{1\times{(T-1)}}\circ I_{\Psi_1 (1:T-1,i)=0}
\end{cases},
\end{eqnarray}
in which $i=1,2,…,n$. In this way, we can get the matrix composed of numerator and denominator of three groups of formulas, (\ref{pjtoi})-(\ref{epsi}). Here, $m:n$ represents getting a new constructed matrix that formed by taking the $m$-th row to $n$-th row or the $m$-th column to $n$-th column of the matrix. It also represents constructing the new vector composed of the $m$-th element to the $n$-th element of the vector. $I_{A=0}$ is the discriminant matrix showing whether the elements in $A$ is 0. If $A(i,j)=0$, then $I_{A=0}(i,j)=1$. The $1_{m\times n}$ matrix represents a matrix in which all elements are 1, this matrix can broadcast the elements in the vector. When writing a program, we also need to change the value of the diagonal marker according to the specific program rules. After this expression is written, we can set the initial value for the three iteration variables, and iterate between (\ref{r1m})-(\ref{rem}) and (\ref{p1m})-(\ref{pem}). We do not end the iteration until the difference between the norm of the three matrices and the previous step is less than a certain value. We have assumed that the process is convergent before iteration. From Fig.\ref{P1} we can see although the symmetrical elements can be vaguely seen on a closer look in $Pt_1$, they are still very inconspicuous. Therefore, this matrix cannot be directly used as an estimate of the adjacency matrix. The final result of the EM algorithm $P_1$ can be approximately regarded as a symmetric matrix, and an adjacency matrix can be generated after transformation. For the calculation of specific single element, we can refer to Appendix \ref{ML} and \ref{vector}.

\subsubsection{A two-step iterative algorithm based on 1-simplex complex}

\begin{figure}[htbp]
    \begin{minipage}[t]{\linewidth}
        \centering
        \includegraphics[width=1\textwidth,trim=0 100 0 130,clip]{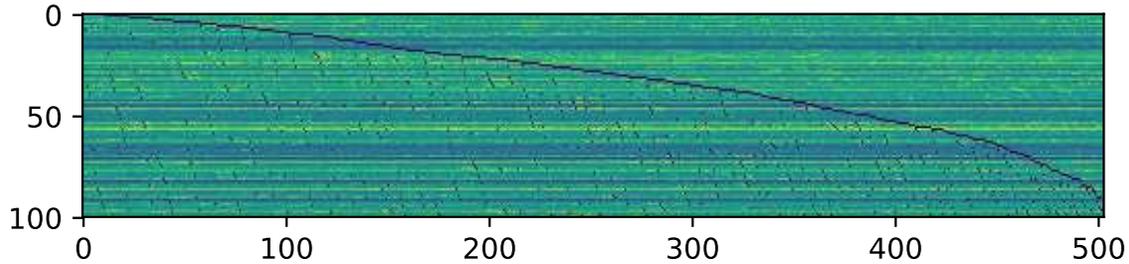}
    \end{minipage}
      \caption{$Pt_2$ matrix generated by the two-step iterative method.}
      \label{Pt2}
\end{figure}

\begin{figure}[htbp]
    \begin{minipage}[t]{\linewidth}
        \centering
        \includegraphics[width=0.9\textwidth,trim=0 0 0 10,clip]{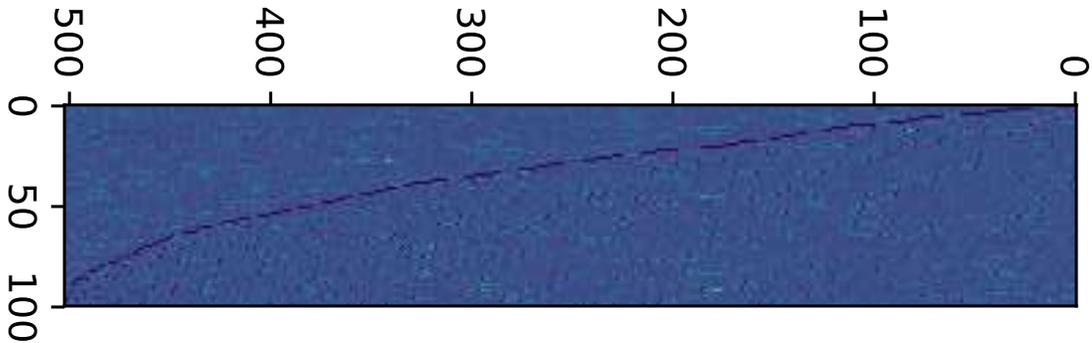}
    \end{minipage}
    \caption{$P_2$ matrix generated by the two-step iterative method.}
    \label{P2}
\end{figure}

If the network is complex and the number of nodes is large, calculating the propagation probability between nodes will lead to a significant increase in the amount of calculation. For example, in a network with $n$ nodes, if the propagation probability of 1-simplex to a single node is calculated respectively, there will be $n(n-1)(n-2)/2$ probability values $P_{jk\to i}$ calculated. Therefore, if we want to add the calculation of the values of $P_{jk\to i}$, we need a simplified algorithm.

At the first step, we need to use the vectorization expression to calculate the reduction of 1-simplex, that is the reconstruction network without 2-simplex. Since every pair of nodes must have connected edges is the necessary condition for the existence of simplex, we can use the reconstructed network to reconstruct 2-simplex. As propagation on 2-simplex is not considered at the first step, we define
\begin{eqnarray}\label{stm}
    S_0^{t_m}=(P_0^T \ast Pt_1)\circ\Psi_1^{t_m},
\end{eqnarray}
The iterative intermediate value and the probability matrix obtained at the first step are expressed as
\begin{eqnarray}\label{r0m}
    R_0^{t_m}=[(1_{n\times 1}\circ\Psi_1^{t_m})\ast(P_0^T \ast Pt_1)]/[(Ep_0+S_0^{t_m})\circ 1_{1\times n}],
\end{eqnarray}
\begin{eqnarray}\label{re0m}
    R_{\varepsilon_0}^{t_m}=Ep_0/[(Ep_0+S_0^{t_m})\circ 1_{1\times n}].
\end{eqnarray}
\begin{eqnarray}\label{p0m}
    P_0(:;i)=(S_{01}^i/S_{02}^i)^T,
\end{eqnarray}
\begin{eqnarray}\label{pe0m}
    \varepsilon_{0i}=(S_{\varepsilon_{01}}^i/S_{\varepsilon_{02}}^i)^T,
\end{eqnarray}
in which
\begin{eqnarray} \label{p0s}
    \displaystyle\begin{cases}
        \displaystyle S_{01}^i=1_{1\times{(T-1)}}\circ\{R_0^{1:(T-1)}(i;:)\ast[(I_{\psi_1(1:T-1,i)=0}\ast\Psi_1(2:T,i))\circ 1_{1\times n}]\}\\
        S_{02}^i=1_{1\times{(T-1)}}\circ\{\Psi_1 (1:T-1,:)\ast[1_{(T-1)\times1}\circ Pt_1(i,:)]\ast[I_{\Psi_1 (1:T-1,i)=0}\circ 1_{1\times  n}]\}
    \end{cases},
\end{eqnarray}
\begin{eqnarray} \label{pe0s}
    \displaystyle\begin{cases}
        \displaystyle S_{\varepsilon_{01}}^i=1_{1\times{(T-1)}}\circ[R_{\varepsilon_0}^{1:(T-1)}(i)\ast I_{\psi_1(1:T-1,i)=0}\ast\Psi_1(2:T,i)]\\
        S_{\varepsilon_{02}}^i=1_{1\times{(T-1)}}\circ I_{\Psi_1 (1:T-1,i)=0}
    \end{cases}.
\end{eqnarray}
Through the iterative operation of (\ref{r0m})-(\ref{pe0m}), we can obtain the probability matrix $P_0=(P_{j\to i})_{n\times n}$, the noise $Ep_0=(\varepsilon_{0i})_{n}$ and the corresponding three-dimensional matrix of the intermediate values. We substitute the results of (\ref{r0m})-(\ref{pe0m}) into the formulas of (\ref{r1m})-(\ref{rem}) and (\ref{p1m})-(\ref{pem}) at the second step of iterative operation. Under this condition, $n_1$ is the number of the network edges determined by the $P_0$ matrix. For a specific node, we only need to find the 2-simple complex containing the node from the three body community derived at the first step. This way can reduce the amount of computation significantly, as it is more efficient than that of traversing all $n(n-1)/2$ node pairs $(i,j)$. Figs.\ref{P1}, \ref{Pt2} and \ref{P2} shows the four matrices generated by our experiment.

In this paper, we use two-step iterative algorithm in the following propagation simulation experiments for different types of networks to ensure the convenience of operation.

\subsection{Confirmation threshold}

This subsection describes how to transform the propagation probability matrices $P_1$ and $P_2$ into adjacency matrices. In 2.1, according to the EM algorithm, we got matrices $P_1$ and $P_2$. The elements $P_{j\to i}$ and $P_{jk\to i}$ in the matrices both satisfy  $P_{j\to i}\ge0$ and $P_{jk\to i}\ge0$. However, there must be errors arised in the calculation, so if we say simplex exist once $P_{j\to i}>0$ or $P_{jk\to i}>0$, inevitable misjudgment of the existence of simplex will be lead by some small probability errors. Therefore, we need to redefine a threshold $\widetilde{P}$ in judging the existence of simplex.

The specific method is as follows. For any node $i$, $\widetilde{P}$ represents the vector $P_1 (:;i)$ with the elements $P_{j\to i}>0$ rearranged from large to small, and $\widetilde{P}_j$ represents the $j$-th probability element of $\widetilde{P}$. We define the node adjacency matrix as $A_{1, n\times n}$, and the adjacency matrix between nodes and 1-simplexes as $A_{2, n_1\times n}$.

In order to find the threshold, we need to find a pair of probabilities with the largest difference in the sorted probabilities. But we must also take into account the different results of subtraction and division. Therefore, we can combine the difference value and the ratio as the basis for evaluation. So we use the formula
\begin{eqnarray} \label{deltap}
    \displaystyle\Delta P(l)=(\widetilde{P}_l-\widetilde{P}_{l+1})\frac{\widetilde{P}_l}{\widetilde{P}_{l+1}}
\end{eqnarray}
to find the value of $l$ that maximize $\Delta P$, defined as
\begin{eqnarray} \label{lhat}
    \displaystyle\hat{l}=\mathop{\arg\max}\limits_{l}\Delta P(l).
\end{eqnarray}
Then we get
\begin{eqnarray} \label{phat}
    \displaystyle\hat{P}_i=\widetilde{P}_{\hat{l}},
\end{eqnarray}
as a threshold for the probabilities in the vector $P(:;i)$. If one of $P_{j\to i}<\hat{P}_i$ or $P_{i\to j}<\hat{P}_j$ holds, it is determined that $i,j$ do not have a connecting edge, or says there is no 1-simplex exists. If $P_{j\to i}\ge\hat{P}_i$ and $P_{i\to j}\ge\hat{P}_j$ are both established, $i$ and $j$ are considered to be connected, then there is a 1-simplex $(i,j)$ and $A_1(i,j)=A_1(j,i)=1$. Similarly, we can calculate $\hat{P}$ under 2-simplicial complex, only when $P_{jk\to i}\ge\hat{P}_i$, $P_{ik\to j}\ge\hat{P}_j$ and $P_{ij\to k}\ge\hat{P}_k$ are all established , 2-simplex $(i,j,k)$ exists. Then $A_2(jk,i)=A_2(ik,j)=A_2(ij,k)=1$. For the remaining nodes or communities without 1-simplex or 2-simplex connections, we define the corresponding elements in the adjacency matrix as 0.

\begin{figure}[htbp]
	\begin{minipage}[c]{0.5\textwidth}
		\centering
		\includegraphics[width=1\textwidth,trim=50 0 60 30,clip]{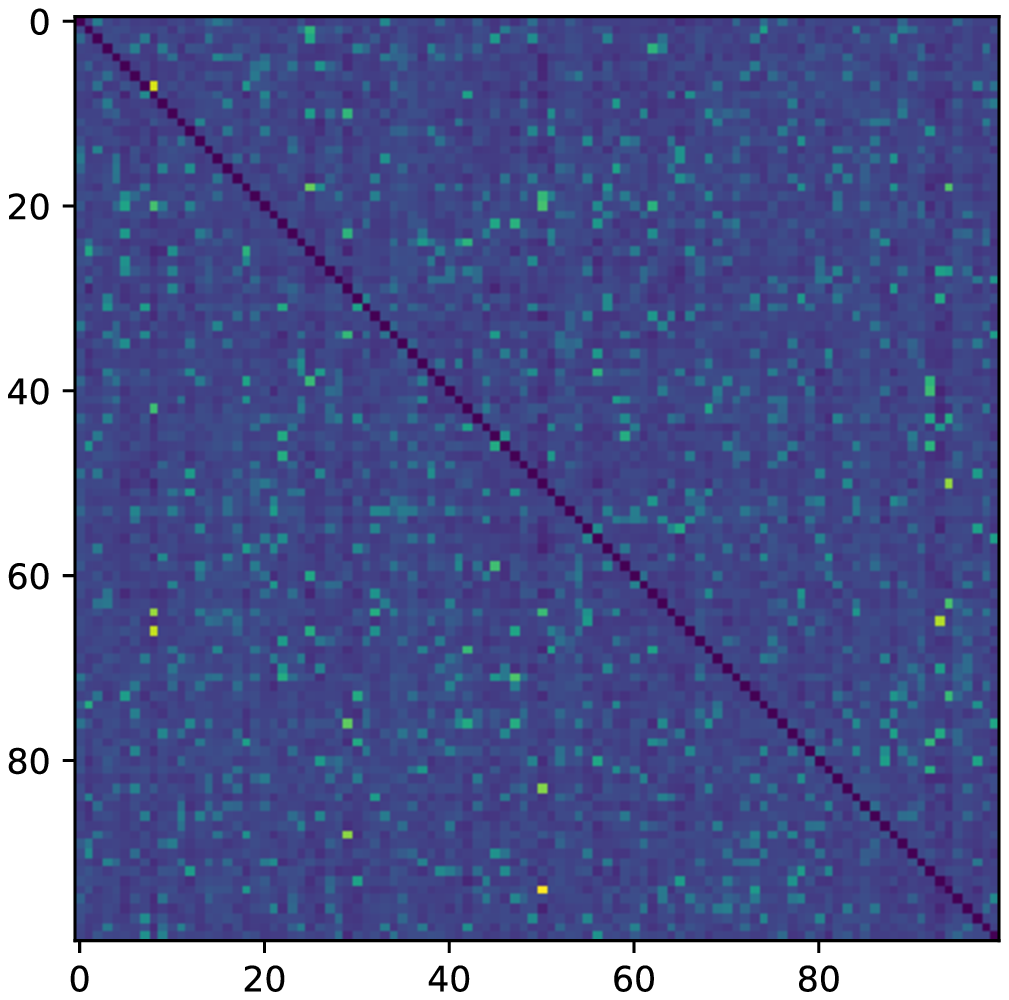}
		\centerline{(a)}
	\end{minipage}
	\begin{minipage}[c]{0.5\textwidth}
		\centering
		\includegraphics[width=1\textwidth,trim=60 0 50 30,clip]{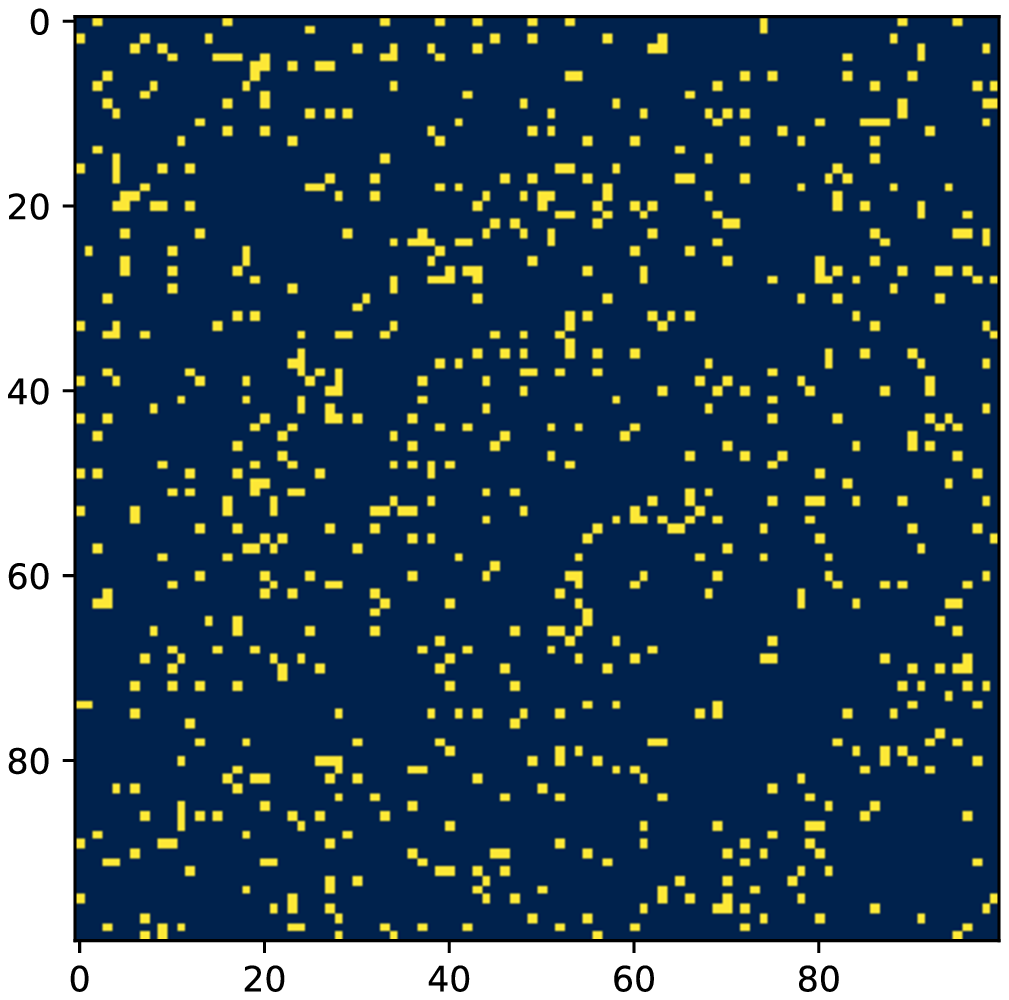}
		\centerline{(b)}
	\end{minipage}
	\caption{(a)$P_1$ of a small-world network. (b)The adjacency matrix $A_1$ generated by transforming the probability matrix $P_1$.}
	\label{adjtrans}
\end{figure}

In order to get an effective critical value, we need to have enough data volume, the larger the data volume, the better the discrimination effect. We generally believe that in propagation experiments, the experimental results are valid only when the number of iterations $T\ge8000$. $P_1$ of a small-world network and the adjacency matrix $A_1$ generated by transforming the probability matrix $P_1$ are shown in Fig(\ref{adjtrans}).

\subsection{Evaluation criteria for effectiveness}
We use $F1$-score to quantify the effectiveness of complex network reconstruction \cite{Powers2020},
\begin{eqnarray} \label{F1}
F1=\frac{2P\times R}{P+R}
\end{eqnarray}
where $P=TP/(TP+FP)$ and $R=TP/(TP+FN)$. $TP, FP, TN$ and $FN$ are the number of true positive, false positive, true negative, and false negative classes, respectively. The larger the $F1$ value, the higher the corresponding accuracy, then the higher the effectiveness of the network reconstruction. $F1=1$ means that the original network structure has been completely reconstructed without errors.

\begin{figure}[htbp]
	\begin{minipage}[c]{0.5\textwidth}
		\centering
		\includegraphics[width=1\textwidth,trim=50 0 60 30,clip]{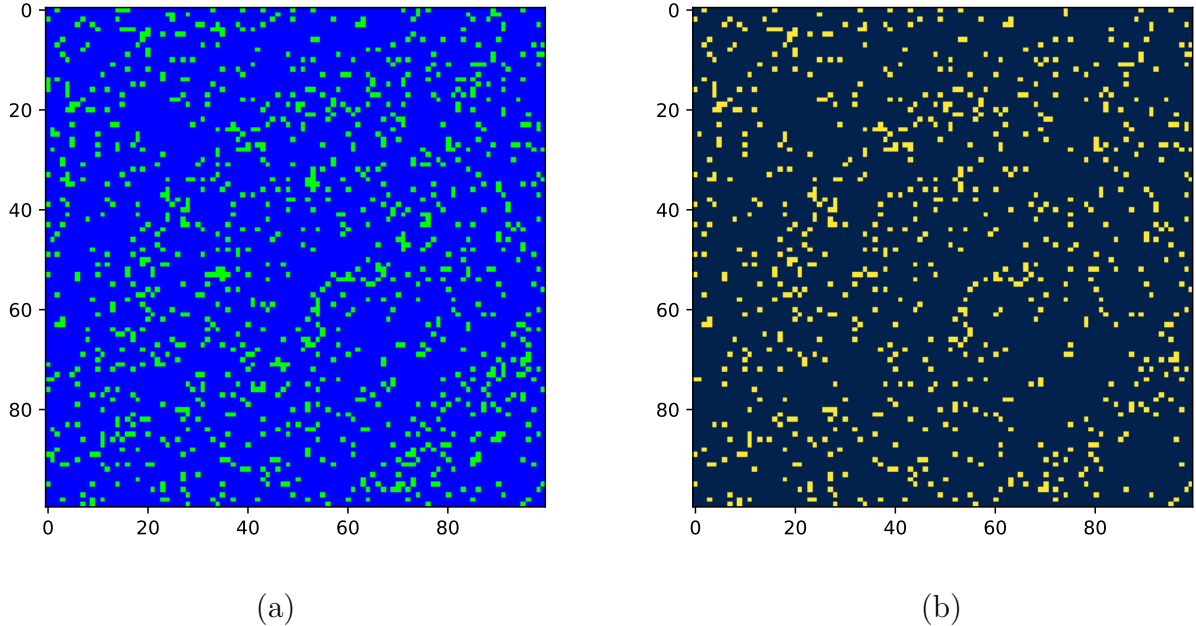}
		\centerline{(a)}
	\end{minipage}
	\begin{minipage}[c]{0.5\textwidth}
		\centering
		\includegraphics[width=1\textwidth,trim=60 0 50 30,clip]{guadj.eps}
		\centerline{(b)}
	\end{minipage}
	\caption{(a)The true adjacency matrix $A_T$. (b)The predicted adjacency matrix $A_1$.}
	\label{f1}
\end{figure}

The true adjacency matrix between network nodes is defined as $A_{T, n\times n}$. $TP$ represents the number of node pairs $(i,j)$ with $A_T(i,j)=A_1(i,j)=1$. $FN$ represents the number of node pairs $(i,j)$ with $A_T(i,j)=1, A_1(i,j)=0$. $FP$ represents thre  number of $A_T(i,j)=0, A_1(i,j)=1$.  $TN$ represents the number of $A_T(i,j)=A_1(i,j)=0$. The comparison of the two matrices is shown in Fig(\ref{f1}). $A_2$ is treated the same way.

\section{Experiment}

\begin{figure}[htbp]
    \begin{minipage}[t]{\linewidth}
        \centering
        \includegraphics[width=0.9\textwidth,trim=0 0 0 0,clip]{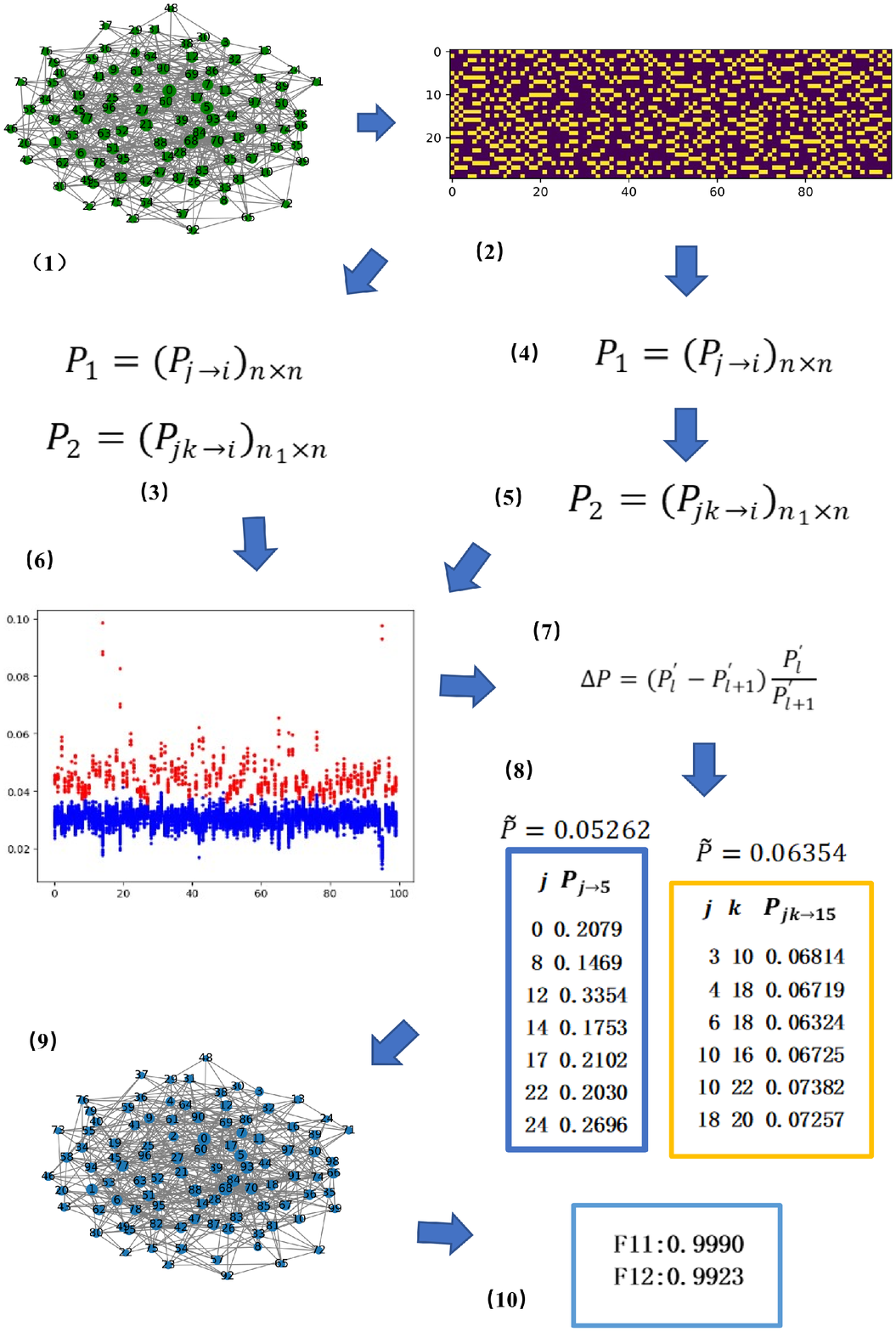}
    \end{minipage}
    \caption{Experiment process}
    \label{experiment}
\end{figure}

In Section 2, we have introduced the steps of complex network reconstruction. The specific process can be summarized in Figure \ref{experiment}. Next, we introduce the specific process of the experiment.

\subsection{Experiment process}

In Fig.\ref{experiment}(1), we specify a network composed of 100 nodes. Edges are randomly generated between nodes. We first use uniform distribution to generate edges. The specific method is as follows. We specify a random probability $p_0$, and each point pair randomly generates a probability $p\sim U(0,1)$ that obeys a uniform distribution. If $p<p_0$, the edge exists, and if $p\ge p_0$, the edge does not exist. In Fig.\ref{experiment} we specify $p_0=0.1$. We have already learned in Section 2 that we can associate the probability of two-body propagation and three-body propagation with the average two-body and three-body degrees of nodes. We specify $\beta_1=\alpha/k_1$, while $\beta_2 =\omega/k_2$, where $\alpha$ and $\omega$ are two predefined parameters, $\alpha<\omega$, and in order to satisfy the basic conditions of probability, we set $\alpha<k_1$ and $\omega<k_2$. In Fig.\ref{experiment}(1), we specify that $\alpha=1.5$ and $\omega=1$. At the same time, the proportion of initial infected nodes is $\rho_0=0.3$, and the probability of infected nodes returning to the susceptible state is $\mu=1$.

After starting the simulation, we can generate the infection state matrix $\Psi_1$ in Fig.\ref{experiment}(2) or Fig.\ref{psi}, and then we can use the one-step iterative algorithm as shown in Fig.\ref{experiment}(3) or the two-step iterative algorithm as shown in Fig.\ref{experiment}(4)-(5). After the above iteration, matrices $P_1=(P_{j\to i})_{n\times n}$ and $P_2=(P_{jk\to i})_{n_1\times n}$ can be generated. According to the formula (\ref{deltap}), (\ref{lhat}) and (\ref{phat}), we can find the critical value of probability, as shown in Fig.\ref{experiment}(6)-(7). Taking Figure \ref{experiment}(8) as an example, for node 5, the threshold of 1-simplex is $\widetilde{P}=0.05262$, so when $P_{j\to5}\ge0.05262$, it means that node $j$ has an edge with node 5, so nodes $j$ and 5 can form a 1-simplex. Similarly, the threshold for 2-simplex judgment of node 15 is $\widetilde{P}=0.06354$, so if $P_{jk\to15}\ge0.06354$, it means that the node pair $(j,k)$ and node 15 can form a 2-simplex. Other nodes follow the same principle. Figure \ref{experiment}(9) is the network reconstructed by the two-step iterative algorithm. The number of iterations in our propagation process satisfies $T=20000>8000$.

Then, according to the two-step iterative method, the first step is to generate a reconstructed 1-simplex network, and the second step is to analyze the network with 2-simplex. According to formula (\ref{F1}), we can know that the network reconstruction accuracy or effectiveness $F1$ of 1-simplex and 2-simplex is 0.9990 and 0.9923, respectively. It can be seen that since the reconstruction of the 2-simplicial complex is carried out on the basis of the reconstruction of the 1-simplicial complex, the reconstruction effectiveness is slightly lower. However, it can be seen that the reconstruction effectiveness or accuracy of 1-simplex and 2-simplex satisfy $F1>0.8$, so it can be concluded that the restoration effect of simplex complex in Fig.\ref{network_example}(a) is well. Generally, we believe that if $F1>0.8$, the network reconstruction meets the standard.

\begin{figure}[htbp]
    \begin{minipage}[c]{0.5\textwidth}
        \centering
        \includegraphics[width=1\textwidth,trim=0 0 0 0,clip]{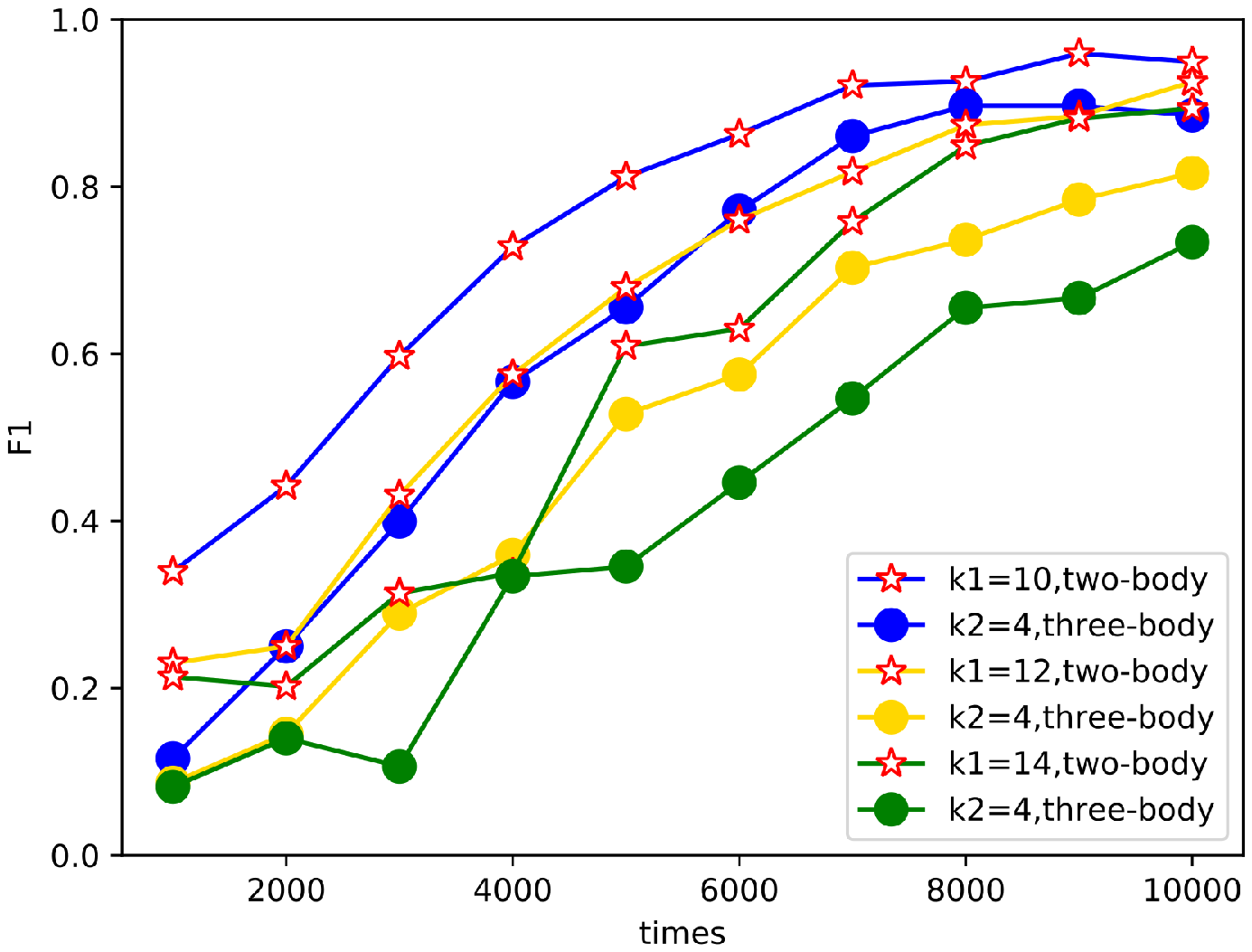}
        \centerline{(a)}
    \end{minipage}
    \begin{minipage}[c]{0.5\textwidth}
        \centering
        \includegraphics[width=1\textwidth,trim=0 0 0 0,clip]{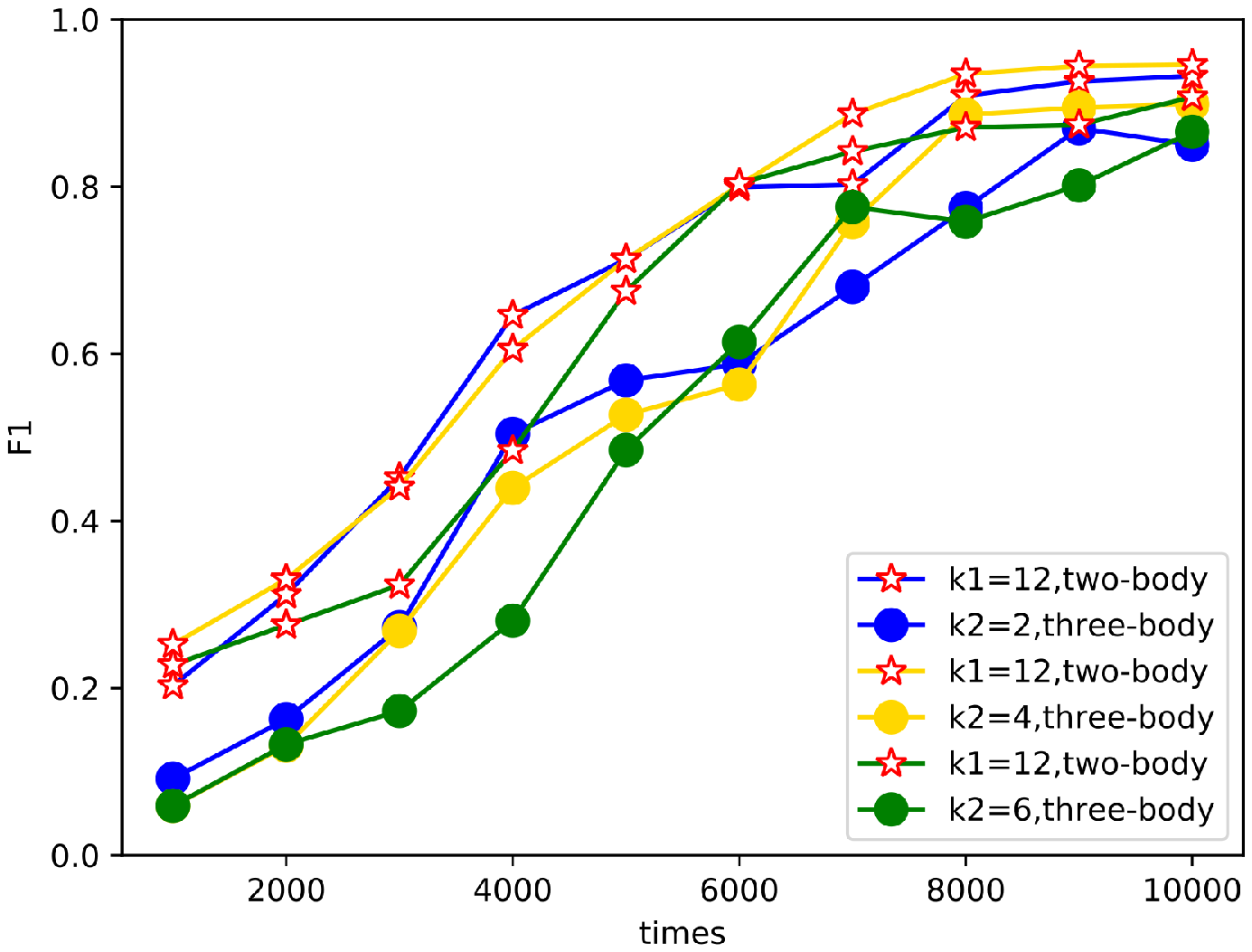}
        \centerline{(b)}
    \end{minipage}
    \caption{The relationship between the $F1$ value and the number of iterations, (a)the change of $F1$ when $k_2$ is fixed and $k_1$ varies; (b)the change of $F1$ when $k_1$ is fixed and $k_2$ varies. It can be seen that no matter what the average degree is, the reconstruction effectiveness will increase with the increase of the number of iterations. The initial parameters of each set of experiments satisfy $\alpha=1.5, \omega=1, \rho_0=0.3$ and $\mu=1$. Since the reconstruction of the 2-simplex complex is based on the reconstruction of the 1-simplex complex, the reconstruction effectiveness is slightly lower.}
    \label{k1k2}
\end{figure}

In order to explore the influence of the reconstruction effectiveness by the number of iterations, we can put the changes of the restoration effectiveness of 1-simplex and 2-simplex with the number of iterations into a unified table and draw several sets of line graphs, as shown in Fig.\ref{k1k2}. The figure shows the relationship between the $F1$ value and the number of iterations $T$ under different average degrees $k_1$ and $k_2$. We specify the starting values $\alpha=1.5, \omega=1, \rho_0=0.3$ and $\mu=1$. Figure \ref{k1k2}(a) means that $k_2=4$ is fixed, and when $k_1$ changes, $F1$ will increase with the increase of the number of iterations. Figure \ref{k1k2}(b) means that $k_1$ is fixed, and when $k_2$ changes, the change trend of $F1$ is the same. It can be seen that, regardless of the average degree, the reconstruction effectiveness or accuracy increases with the number of iterations. Therefore, in the following analysis, the difference in the average degree does not affect the variation law of $F1$, $T$ and other values. It can be seen from Fig.\ref{k1k2} that no matter what the average degree is, the reconstruction effectiveness will increase with the increase of the number of iterations. In other words, the error in the mean degree does not affect the overall trend.

Since the reconstruction of the 2-simplex complex in the two-step iterative method is based on the reconstruction of the 1-simplex complex, the effectiveness of the reconstruction is slightly lower, which can be verified in the graph. When $T\ge8000$, the reconstruction effectiveness of 1-simplicial complex can satisfy $F1>0.8$, and 2-simplicial complex requires higher iterations under certain conditions. In the following content, we conduct specific analysis on different types of networks.

\begin{figure}[htbp]
    \begin{minipage}[c]{0.5\textwidth}
        \centering
        \includegraphics[width=1\textwidth,trim=30 30 30 30,clip]{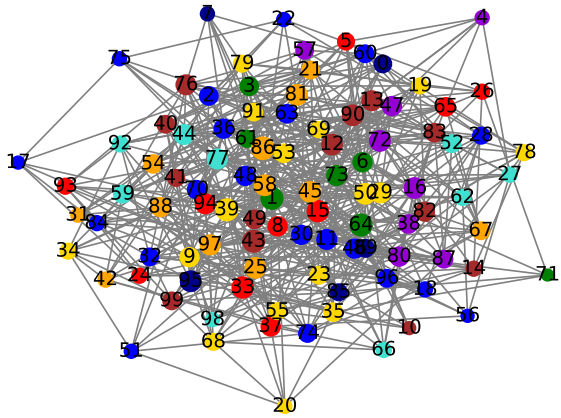}
        \centerline{(a)}
    \end{minipage}
    \begin{minipage}[c]{0.5\textwidth}
        \centering
        \includegraphics[width=1\textwidth,trim=30 30 30 30,clip]{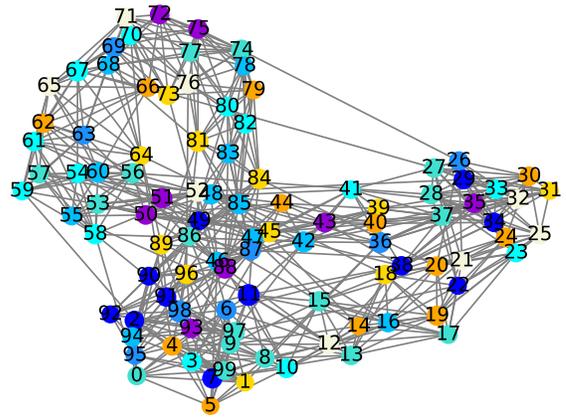}
        \centerline{(b)}
    \end{minipage}
    \begin{minipage}[c]{0.5\textwidth}
        \centering
        \includegraphics[width=1\textwidth,trim=30 30 30 30,clip]{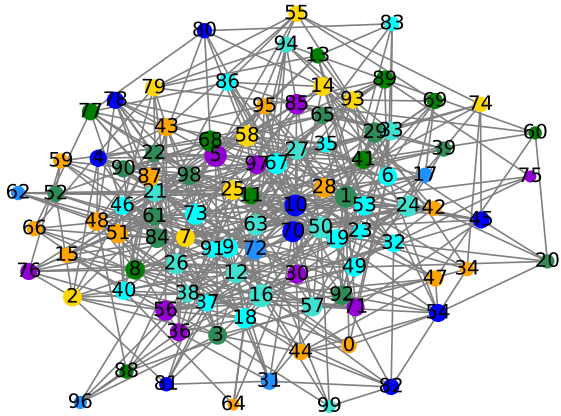}
        \centerline{(c)}
    \end{minipage}
    \begin{minipage}[c]{0.5\textwidth}
    \centering
    \includegraphics[width=1\textwidth,trim=30 30 30 30,clip]{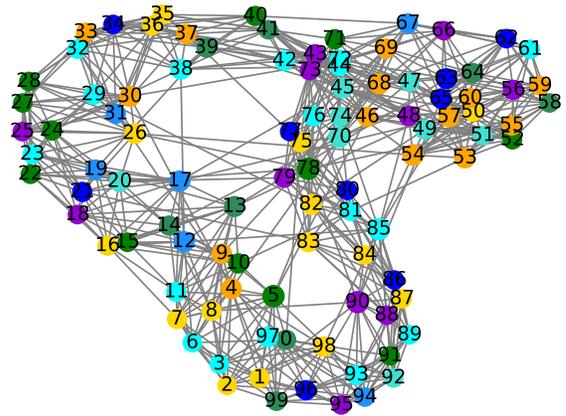}
    \centerline{(d)}
    \end{minipage}
    \caption{The four networks with the best experimental results, (a)Gaussian network, (b)Watts network, (c)GNP network and (d)Newman Watts network. The probability distribution of the Gaussian network is a continuous Gaussian distribution. The Watts network is also called the WS small world network. The GNP network returns a GNP random graph, and no edges are deleted in the Newman Watts network.}
    \label{graph}
\end{figure}

In addition to the binomial distribution network, we also analyzed the reconstruction of various networks, four of which also performed well. These four networks are also the best experimental networks. As shown in Fig.\ref{graph}, the four networks are (a)Gaussian network, (b)Watts network, (c)GNP network and (d)Newman Watts network. Gaussian network is a probabilistic graphical model. For ordinary probabilistic graphical models, the probability distribution of random variables is discrete, while the probability distribution of Gaussian network is continuous Gaussian distribution \cite{Rader2005}. Watts network is also called WS small-world network. The method to generate it is to first create a ring on $n$ nodes and connect each node in the ring to its $k$ nearest neighbors, or $k-1$ if $k$ is odd. Then create the entire graph network by replacing some edges. For example, each edge $(u, v)$ in the underlying $n$-ring with $k$ nearest neighbors, the edge is replaced by a new edge $(u, w)$ with probability $p$ , where $w$ uniformly randomly selects existing nodes \cite{Watts1998, Kossinets2006}. The GNP network returns a GNP random graph, also known as an Erdős-Rényi graph \cite{Seshadhri2012}. The GNP graph algorithm chooses each of the $[n(n-1)]/2$ possible edges with probability $p$. The Newman Watts network is generated in a similar way to the Watts network, except that no edges are deleted \cite{Newman1999, Fu2007}. Compared with other networks, the Newman Watts network and the Watts network are characterized in that they are a class of networks with shorter average path lengths, and it can be said that the entire network structure is flatter. In the subsection, we conduct simulation experiments on four networks respectively. We mainly study the correlation of reconstruction effectiveness on four networks with two different hyperparameters. The effect of other types of network experiments is not obvious, and we put the network reconstruction effectiveness at 10000 iterations in Table \ref{tab8}. These networks all have 100 nodes, and the average degrees $k_1$ are in the interval $9<k_1<21$.

\subsection{The relationship between the effectiveness of network reconstruction and the number of iterations}

\begin{figure}[htbp]
    \begin{minipage}[c]{0.5\textwidth}
        \centering
        \includegraphics[width=1\textwidth,trim=20 0 30 30,clip]{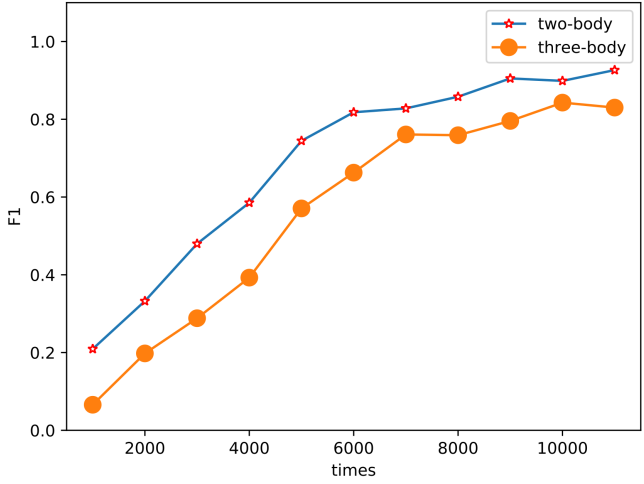}
        \centerline{(a)}
    \end{minipage}
    \begin{minipage}[c]{0.5\textwidth}
        \centering
        \includegraphics[width=1\textwidth,trim=20 0 30 30,clip]{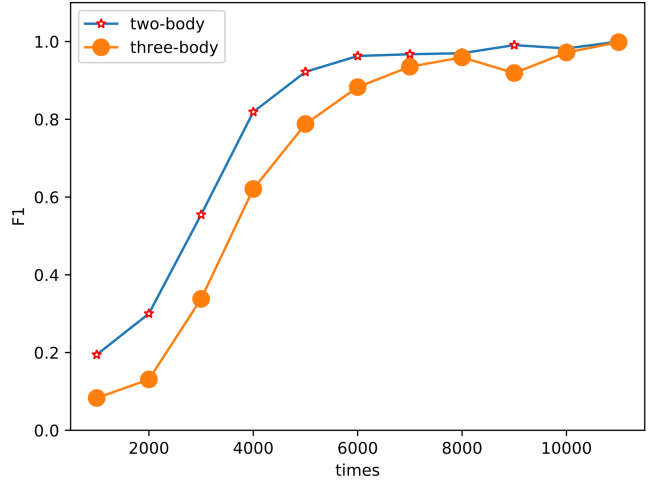}
        \centerline{(b)}
    \end{minipage}
    \begin{minipage}[c]{0.5\textwidth}
        \centering
        \includegraphics[width=1\textwidth,trim=20 0 30 30,clip]{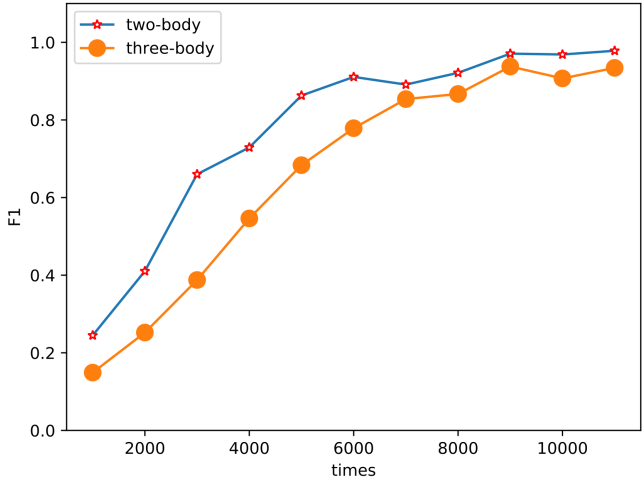}
        \centerline{(c)}
    \end{minipage}
    \begin{minipage}[c]{0.5\textwidth}
        \centering
        \includegraphics[width=1\textwidth,trim=20 0 30 30,clip]{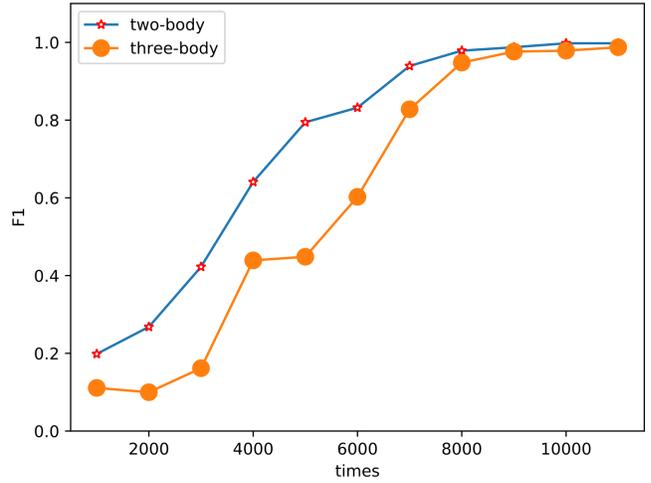}
        \centerline{(d)}
    \end{minipage}
    \caption{Reconstruction effectiveness of (a) Gaussian network, (b) Watts network, (c) GNP network and (d) Newman Watts network as a function of the number of iterations. Starting values $\alpha=1.5, \omega=1, \rho_0=0.3$ and $\mu=1$. When $T\ge8000$, the reconstruction effectiveness of simple complex can satisfy $F1>0.8$ and tend to be stable. While Newman Watts network and Watts network outperform other networks.}
\label{graph_T}
\end{figure}

What we need to do in this section is to analyze the variation law of the reconstruction effectiveness $F1$ with the maximum number of iterations $T$ on the above four networks. We select the network in Fig.\ref{graph} and input it into the process (1) of Fig.\ref{experiment}. At the same time, when performing process (2), we modify the maximum number of iterations $T$. After the data is generated, the change in reconstruction effectiveness $F1$ with $T$ is visualized, as shown in Fig.\ref{graph_T}. Figure \ref{graph_T} reflects the variation of the reconstruction effectiveness (or accuracy) of (a) Gaussian network, (b) Watts network, (c) GNP network and (d) Newman Watts network with the number of iterations $T$.

For network experiments on four types of networks, we all specify the starting values $\alpha=1.5, \omega=1, \rho_0=0.3$ and $\mu=1$. From the images and data, we can judge that the reconstruction effectiveness F1 increases with the increase of the number of iterations T. Among them, when $T\ge8000$, the reconstruction effectiveness of simplicial complex satisfies $F1>0.8$ and tends to be stable. Since the reconstruction of the 2-simplicial complex is carried out on the basis of the reconstruction of the 1-simplicial complex, the reconstruction effectiveness is slightly lower. But as long as the amount of data is large enough, the standard $F1>0.8$ can be reached.

By observing the images, we can find that for Watts network and Newman Watts network, when $T\to10000$, the reconstruction effectiveness of two-body and three-body structures in 1-simplex and 2-simplex constantly tends to 1.0. However, there is an obvious gap between the Gaussian network and the GNP network, and the reason why the gap is caused here will be analyzed in Section 4. Finding a network suitable for the reconstruction method and exploring how to solve other types of networks with poor restoration may be a future research direction.

\subsection{The relationship between the effectiveness of network reconstruction and the perturbation during propagation}

\begin{figure}[htbp]
    \begin{minipage}[c]{0.5\textwidth}
        \centering
        \includegraphics[width=1\textwidth,trim=20 0 30 30,clip]{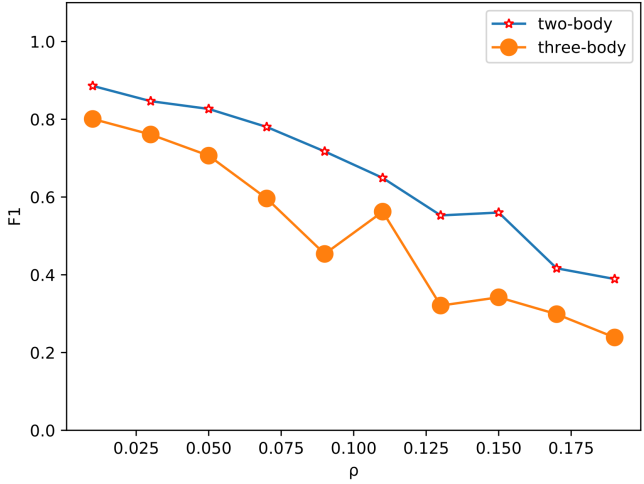}
        \centerline{(a)}
    \end{minipage}
    \begin{minipage}[c]{0.5\textwidth}
        \centering
        \includegraphics[width=1\textwidth,trim=20 0 30 30,clip]{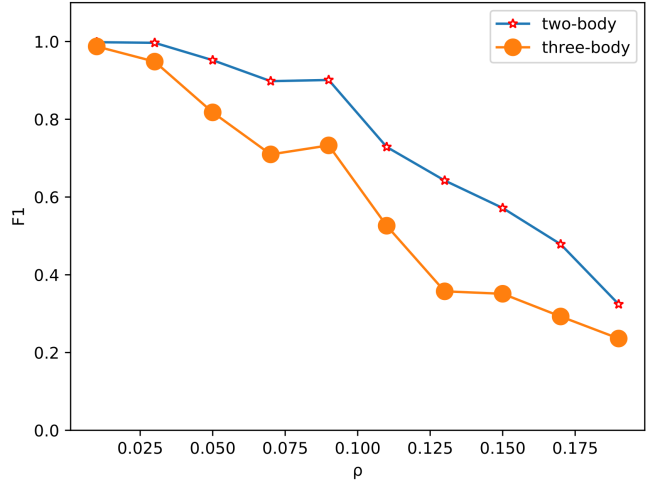}
        \centerline{(b)}
    \end{minipage}
    \begin{minipage}[c]{0.5\textwidth}
        \centering
        \includegraphics[width=1\textwidth,trim=20 0 30 30,clip]{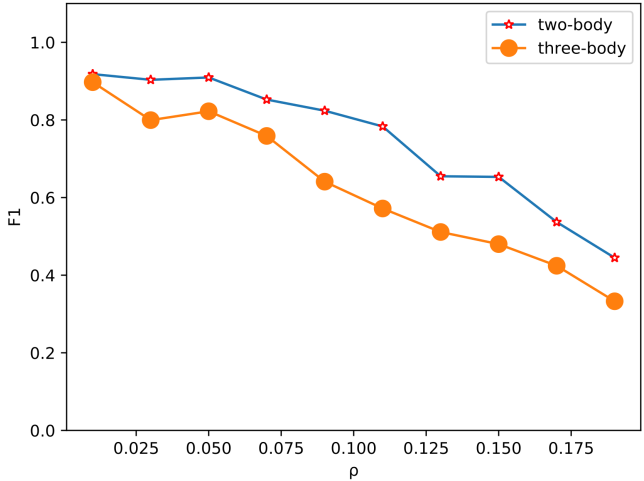}
        \centerline{(c)}
    \end{minipage}
    \begin{minipage}[c]{0.5\textwidth}
        \centering
        \includegraphics[width=1\textwidth,trim=20 0 30 30,clip]{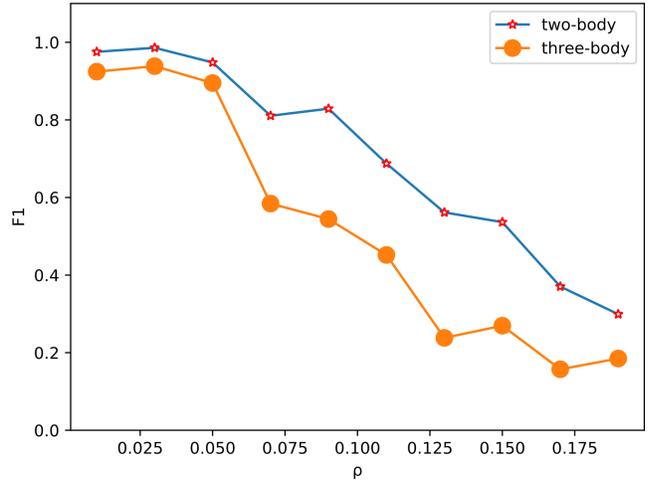}
        \centerline{(d)}
    \end{minipage}
    \caption{Reconstruction effectiveness of (a) Gaussian network, (b) Watts network, (c) GNP network and (d) Newman Watts network as a function of perturbation rate $\rho$. For the four types of networks, we specify the starting values $\alpha=1.5, \omega=1, \rho_0=0.3$ and $\mu=1$. The number of iterations $T=10000$. We can see that in the case of less perturbation, the reconstruction effectiveness can be kept well, $F1>0.8$. However, if $\rho$ is too large, the reconstruction effectiveness is greatly affected, and even the phenomenon of $F1<0.2$ can appear.}
    \label{graph_rho}
\end{figure}

What we need to do in this subsection is to analyze the variation law of the effectiveness $F1$ with the iterative disturbance rate $\rho$ on the above four networks. We select the network in Fig.\ref{graph} and input it into the process (1) of Fig.\ref{experiment}. When performing process (2), we will introduce the perturbation rate $\rho$, and finally visualize the change of effectiveness with $\rho$, as shown in Fig.\ref{graph_rho}. Figure \ref{graph_rho} reflects the reconstruction effectiveness of (a) Gaussian network, (b) Watts network, (c) GNP network and (d) Newman Watts network as a function of perturbation rate $\rho$. For network experiments on four types of networks, we all specify the starting values $\alpha=1.5, \omega=1, \rho_0=0.3$ and $\mu=1$.

The definition of the perturbation rate $\rho$ here means that in the propagation experiment, after each iteration, each node has the probability of $\rho$ to change the infection state, such as $\psi_i^t=1$, then the node has the probability of $\rho$ to become $\psi_i^t =0$. Vice versa, when $\psi_i^t=0$, the node has the probability of $\rho$ to become $\psi_i^t =1$. If there is such a disturbance, the network reconstruction must be affected. For the specific impact, we also use the reconstruction effectiveness as a reference.

Through the data and images, we can see that in the case of less perturbation, $\rho<0.05$, the reduction degree can be maintained well, $F1>0.8$. However, if $\rho$ is too large, the reduction degree will be greatly affected, and even the phenomenon of $F1<0.2$ will appear. The relative size of the perturbation is closely related to the network complexity. All our experiments use a network of 100 nodes, so the perturbation is already very large when $\rho=0.2$. On the whole, the larger the perturbation $\rho$, the lower the network reconstruction effectiveness $F1$. We also test the perturbative effects on networks of 30 and 200 nodes. For a network of 30 nodes, when $\rho=0.1$, $F1<0.2$ has appeared. For a network of 200 nodes, $F1<0.2$ only occurs when $\rho=0.3$. Similar phenomena can be observed from real networks with more nodes \cite{Wang2022}.

Therefore, we conclude that in order to ensure a good network reconstruction when the disturbance exists, we can reduce the disturbance rate or increase the size of the network to reduce the impact of its disturbance.

\section{Analysis of experiments and network types}

In Section 3, by observing the images after data visualization, we can find that for Watts network and Newman Watts network, when $T\to10000$, the reconstruction effectiveness of 1-simplex and 2-simplex tends to 1.0. However, there is an obvious gap between the effectiveness of Gaussian network and the GNP network with $F1=1.0$. Even if $T=10000$, the reconstruction effectiveness is still far from 1.0. We will analyze how the gap is caused in this section.

We search for suitable network types for the iterative algorithm in this paper, and present three methods to improve the efficiency of reconstruction.

\subsection{The relationship between reconstruction effectiveness and network types}

\begin{table}[htbp]
    \setlength{\tabcolsep}{1mm}
    \caption{Reconstruction effectiveness and related data of different types of networks with $n=100$ nodes when the number of iterations $T = 10000$.}
    \label{tab8}
    \centering
    \begin{tabular}{l|ll|ll|ll|ll}\hline
        Network Type   &  $k_1$      & $k_2$    &   $F1_1$     & $F1_2$ &$\sigma^2$  & $\sigma$& $MI$ &$BC$  \\\hline
        Watts Network & 12.00&  20.25&  0.9873& 0.9734& 1.000 &1.000& 0.06886&0.01567  \\
        Newman Watts Network  & 13.06&  29.01&  0.9836  &0.9543 &1.075  &1.037&0.06799& 0.01435 \\
        Binomial Network & 9.700&   2.340&  0.9152& 0.7783&8.243 &  2.871 &0.1739&0.01278 \\
        GNP Network & 10.28 &3.060& 0.9041& 0.7948&10.14&   3.184&0.1571&0.01190 \\
        Gaussian Network & 11.10    &4.050& 0.9184& 0.7729&9.321&   3.053&0.1645&0.01285 \\
        Powerlaw Cluster Network & 17.76&   29.13&  0.5232& 0.1814&77.40 &  8.798& 0.2439& 0.008767  \\
        Barabasi Albert Network & 18.00&    34.56&  0.4721& 0.1506&83.39&  9.132&0.2623& 0.008812 \\\hline
    \end{tabular}
\end{table}

The reconstruction effectiveness and related data of different types of networks with 100 nodes are given in Table \ref{tab8} when $T=10000$. In Table \ref{tab8}, the two-body community degree $k_1$ is the degree of the node we specify in the ordinary network, which means, the number of adjacent nodes that have edges with the node. The three-body community degree $k_2$ is the number of valid 2-simplexes containing the node, which means, the number of triangular communities that can form 2-simplexes. Due to the different network structures, the generated communities that meet the conditions are also different. $F1_1$ and $F1_2$ respectively represent the reconstruction effectiveness of the network when $T=10000$. Powerlaw cluster networks are networks produced by Holme and Kim's algorithm with a power-law degree distribution and a growth graph that approximates average clustering \cite{AbouRjeili2006}. Barabasi Albert network refers to network generation in which a graph of $n$ nodes grows by appending new nodes (each with $m$ edges) that preferentially append to existing nodes with higher order \cite{Albert2002}. Many related theories of random networks \cite{Cameron1997, Newman2002} can already be implemented in Python.

Because we mainly rely on the number of nodes and the distribution of the degrees of nodes when we generate a random network, we find that it is difficult for the degree variance of network nodes to change significantly under the two premises that the same type of network has the same number of nodes as well as the average degree is in the same order of magnitude in the process of network generation. Therefore, once the variance of the degree has a large gap across the order of magnitude, there is a high probability that the type of network structure is different. Since degree variance is hard to adjust and random network types are limited, the relationship of this variable to other variables is rarely mentioned. We compare typical networks in Networkx language package of Python in Table \ref{tab8}. Due to the large difference in degree variance of different networks, for the convenience of visualization, we take the standard deviation of the degree to narrow the gap.  When generating seven kinds of networks in the experiment, we set all expectations of the number of neighbor nodes to 15, so that the average degree of the generated network is about 15.  Though there is no obvious distribution of the network average degree since the number of network types that can be generated in the Networkx language package does not satisfy the law of large numbers, we can confirm that the average degree is all within 20 and in the same order of magnitude. We can also know that slight changes in the degree of the same type of network do not affect the overall trend from Fig.\ref{k1k2}.

\begin{figure}[htbp]
    \begin{minipage}[c]{0.5\textwidth}
        \centering
        \includegraphics[width=1\textwidth,trim=20 0 30 30,clip]{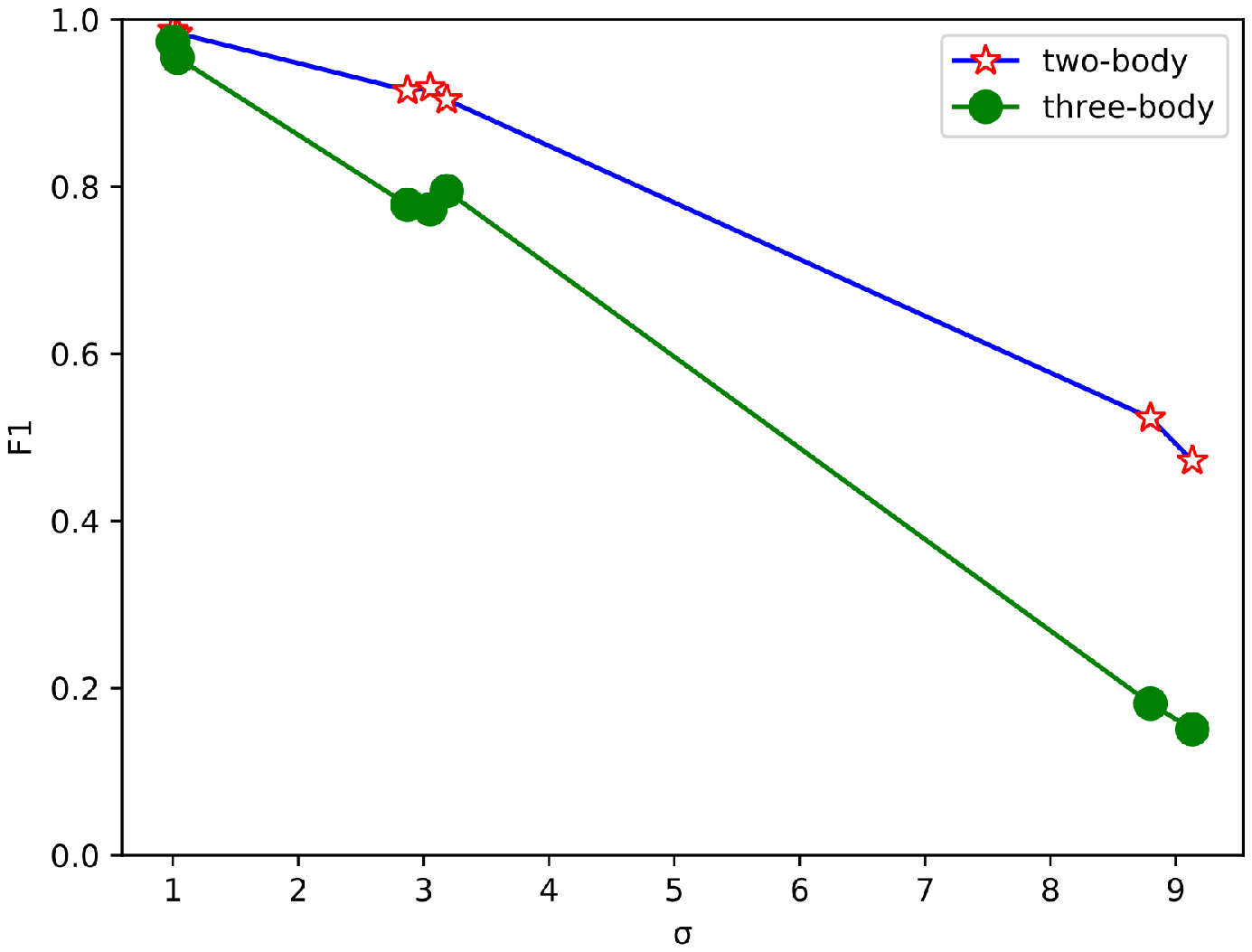}
        \centerline{(a)}
    \end{minipage}
    \begin{minipage}[c]{0.5\textwidth}
        \centering
        \includegraphics[width=1\textwidth,trim=20 0 30 30,clip]{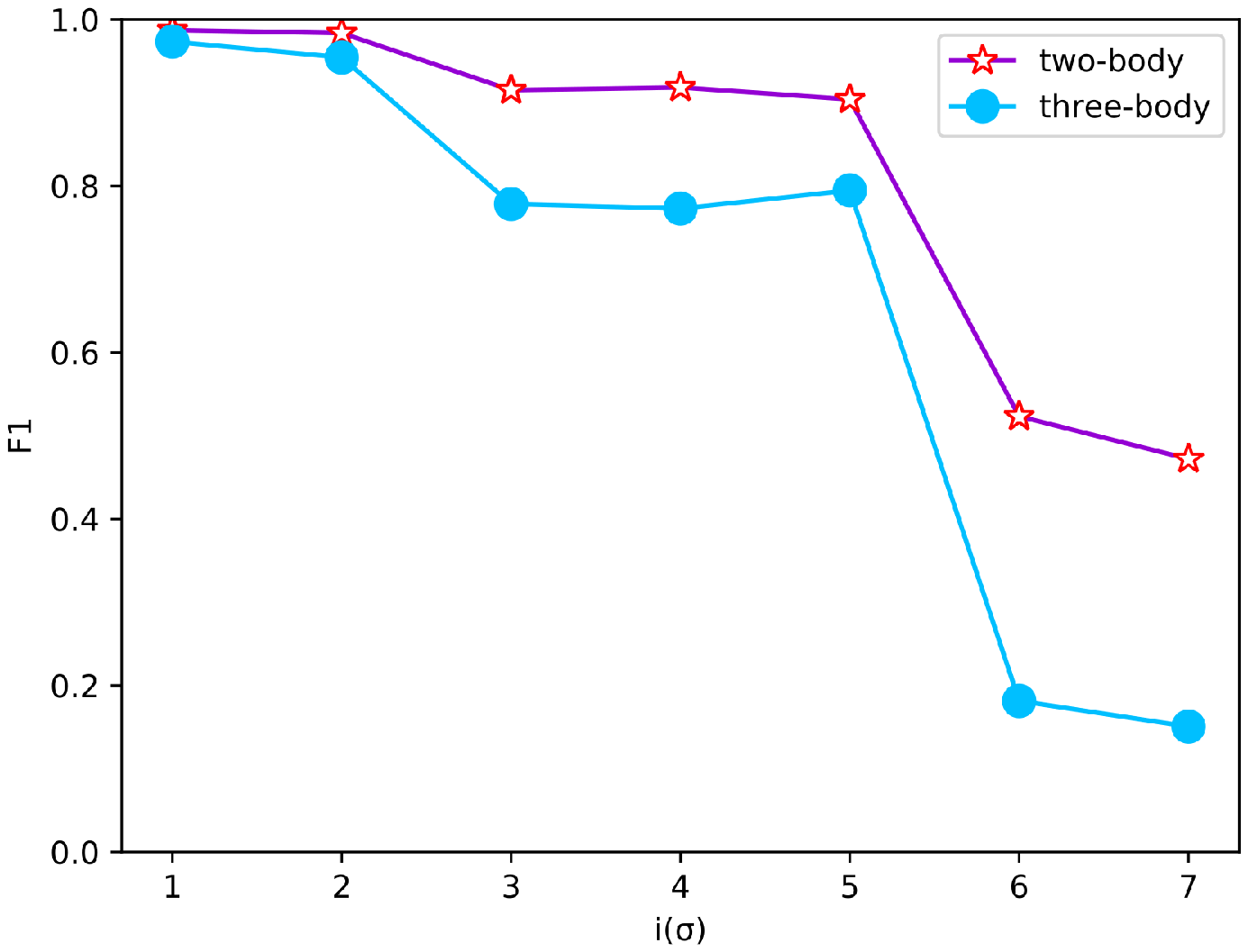}
        \centerline{(b)}
    \end{minipage}
    \caption{(a)The standard deviation $\sigma$ of the seven networks and their corresponding reconstruction effectiveness $F1_1$ and $F1_2$. (b)The order statistics $i(\sigma)$ of the order quantities after $\sigma$ is arranged from small to large and their corresponding reconstruction effectiveness $F1_1$ and $F1_2$.}
    \label{sig}
\end{figure}

We put the standard deviation $\sigma$ of the seven networks and their corresponding reconstruction effectiveness $F1_1$ and $F1_2$ into the chart, as shown in Fig.\ref{sig}(a). Similarly, we put the order statistics $i(\sigma)$ which is the order quantities after $\sigma$ is arranged from small to large and their corresponding reconstruction effectiveness $F1_1$ and $F1_2$ into the chart, as shown in Fig.\ref{sig}(b). It is found that when the number of network nodes is constant and the average degree is in the same order of magnitude, the general trend is the larger the degree standard deviation $\sigma$ of the network, the lower the reconstruction effectiveness $F1_1$ and $F1_2$ of the network. In order to reconstruct the network effectively, it is necessary to ensure the degree variance is not too large, that means, to ensure the difference between the degrees of the nodes is not too large.

The difference of degree variance or degree standard deviation can be regarded as an index to judge the difference of network structure when the number of nodes in a network is the same and the average degree is in the same order of magnitude. From the three networks in Figs.\ref{network_example}(a), \ref{graph}(a) and \ref{graph}(c), we can see that the three networks cannot be distinguished directly, and the degree variances of the three are also very close, so the three network structures are similar. In fact, the binomial network and the GNP network are almost the same types of networks generated using two different languages in Python. However, there are obvious differences in structure between the above two networks and the two small-world networks. This is also the reason why their reconstruction effectiveness is different under the same number of iterations. Although the Newman Watts network and the Watts network are slightly different in the way of generating, they both belong to small-world networks. The above analyzes are consistent with the difference in degree variance. We can also argue that the larger the degree variance, the more the structure differs from the small-world network. 

\begin{figure}[htbp]
	\begin{minipage}[c]{0.5\textwidth}
		\centering
		\includegraphics[width=1\textwidth,trim=20 0 30 30,clip]{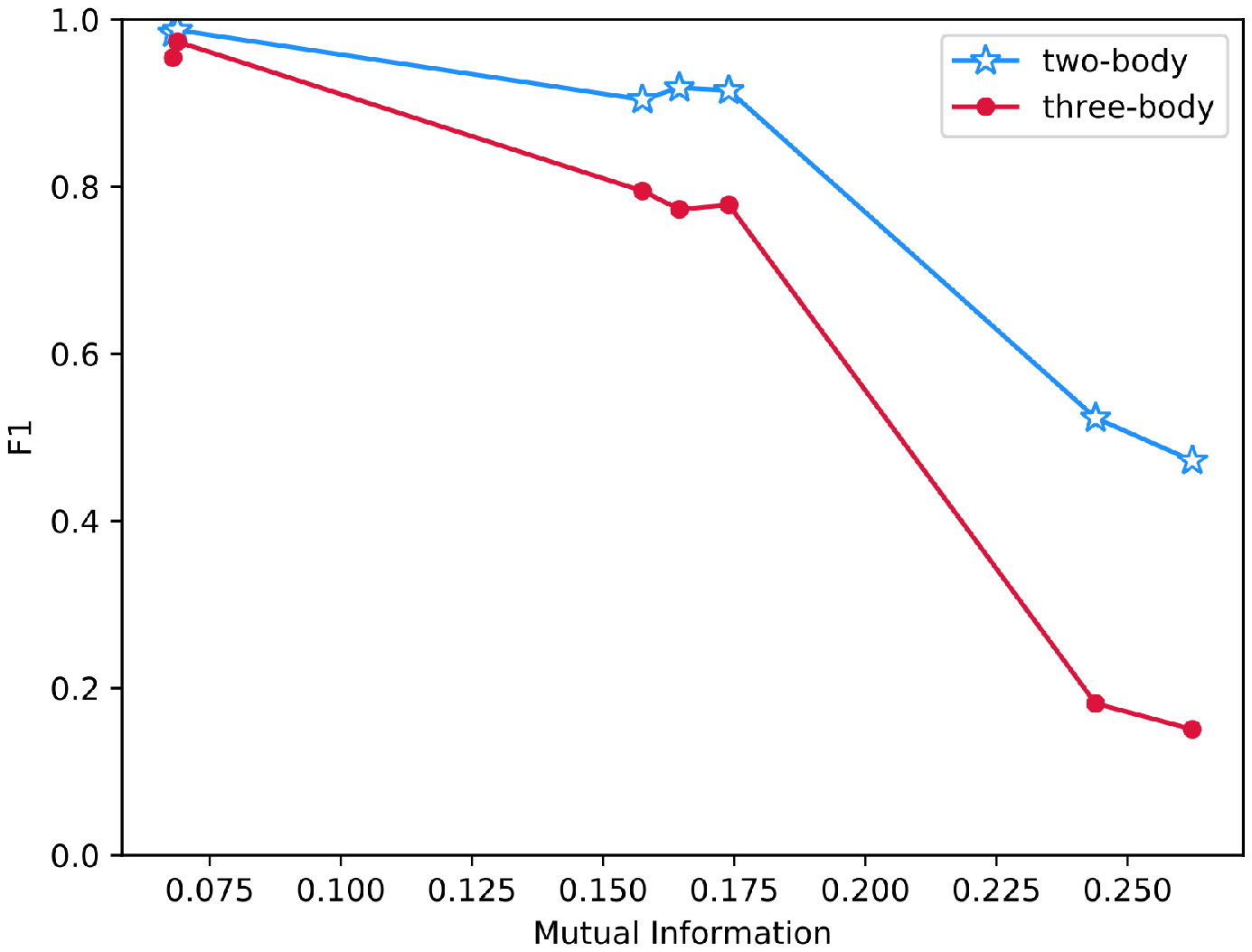}
		\centerline{(a)}
	\end{minipage}
	\begin{minipage}[c]{0.5\textwidth}
		\centering
		\includegraphics[width=1\textwidth,trim=20 0 30 30,clip]{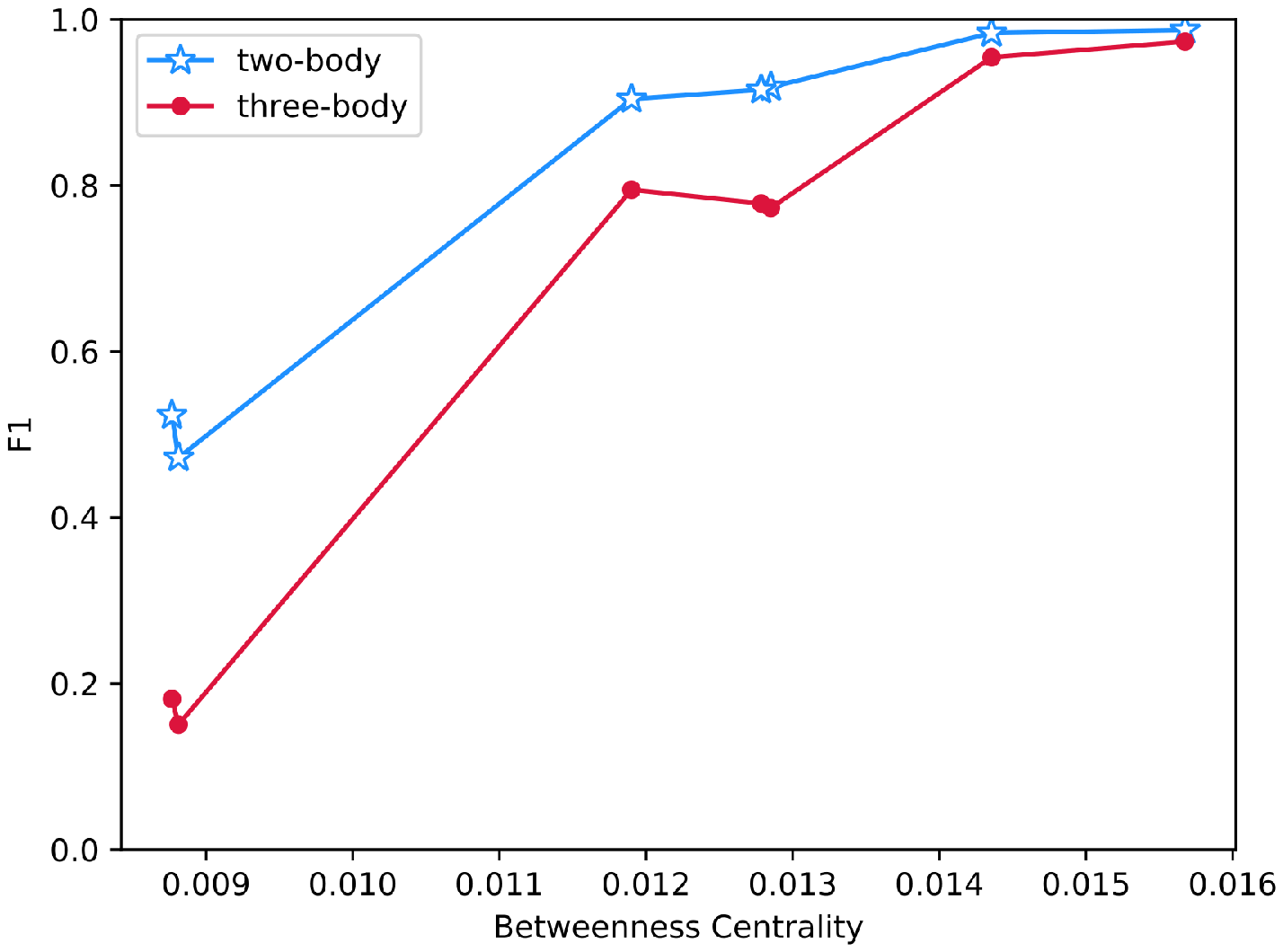}
		\centerline{(b)}
	\end{minipage}
	\caption{(a)The betweenness centrality of the seven networks and their corresponding reconstruction effectiveness $F1_1$ and $F1_2$. (b)The mutual information of the degree distribution of the seven networks and their corresponding reconstruction effectiveness $F1_1$ and $F1_2$.}
	\label{stru}
\end{figure}

In order to test the rationality, we take two other indicators. One indicator is betweenness centrality (BC) \cite{Newman2005}. Compute the shortest-path betweenness centrality for nodes in each network. Another indicator is mutual information (MI) \cite{Maes1996} of degree distribution. We calculate the mutual information of each network degree distribution and small-world network degree distribution. The experimental data can be found in Table.\ref{tab8}, and the visual data can refer to Fig.\ref{stru}. It can be seen that the overall trend is that the reconstruction effectiveness decreases with the increase of mutual information, and increases with the increase of betweenness centrality. Small world networks' mutual information is the smallest and betweenness centrality is the largest. It can be seen that the greater the gap between the network structure and the small world network structure, the worse the reconstruction effect.

In the following content, we mainly analyze the small-world network.

\subsection{Small-world network reconstruction}

In this paper, we can see that the small-world network is the most suitable type of network for the experimental method that the network reconstruction based on binary time series and EM algorithm. This just shows that it is also feasible to restore a information dissemination network between people based on the SIS model in reality. A small-world network is a type of mathematical graph in which most nodes are not adjacent to each other, but most nodes can be reached from any other point in a few steps \cite{Newman1999, Watts1998}.

The actual social, ecological and other networks are all small-world networks. In such a system, the information transmission speed is fast, and a small change of a few connections can drastically change the performance of the network, such as adjusting the existing of nodes or edges in the network. Introduced into reality, if each node in a small-world network represents a person, and the existence of network edges means those people know each other, the small-world network can reflect the phenomenon that strangers are connected by a person they all know. There are many relevant real models, such as networks that reflect people's face-to-face interactions \cite{Stehle2011, Genois2018}, community structure \cite{Karrer2011}, and some cases in schools \cite{Mastrandrea2015}.

The greater the degree variance of the network, the greater the difference between the network and the small-world network structure. In reality, in interpersonal relationships, we can make social networks closer to small-world networks or optimize experimental methods to obtain a model that allows network reconfiguration to proceed smoothly.

\subsection{Methods of improvement}

From Section 3, we know that we can increase the number of iterations or reduce the perturbation rate to improve the effectiveness of the network reconstruction based on binary time series and EM algorithm in the experiment. But in reality, it is difficult to increase the maximum number of iterations in many cases due to the large demand for data and the scarcity of data itself. At the same time, the number of iterations is reflected in the form of time series time points, so the time condition is also a big limitation. Therefore, if we need to restore the relationship network behind data based on infection and rumor spread, we must pursue the data with the most effective results in a limited time. What we can do more is to optimize samples and algorithms. Since the reconstruction efficiency of the small-world network is the largest, we can make the experimental individual relation network closer to the small-world network. We summarize three methods that can make the reconstruction method in this paper achieve higher network reconstruction effectiveness within the same time and iterations.

The specific methods are as follows:

\begin{enumerate}[(1)]
    \item Decentralization.

     Decentralization is diluting the hierarchy and reducing the so-called upper and lower boundaries. Considering only two-body and three-body propagation, if there is obvious centralization, the efficiency of information propagation will be reduced, because the information of the central node, or the node with a larger degree, can only be propagated nearby. It will be difficult for non-central nodes, or the nodes with smaller degrees, to obtain an effective number of infected time nodes within the specified number of iterations. Then the connection between the node and the adjacent node will be misjudged. In order to enable each node to obtain valid data information, it is necessary to ensure that the probability of the existence of edges between nodes and other nodes is similar, which means there should not be too many nodes that are too central or too marginal. This approach, also known as flattening, is typified by the Watts network, which enables smoother propagation of experiments and clearer feedback from experiments. This is something to be aware of when looking for real networks.

    \item Extend network scale.

    That is to obtain a sufficiently large number of individual samples, which ensures that the network has a large enough scale to reduce the impact of individual nodes with too large or too small degrees. When the scale of the network is large enough, the information dissemination range of the entire network will be larger. Even if there are points with a large degree, the proportion of the nodes themselves will be reduced under the influence of the scale. Even around a relatively centralized node, if the information dissemination is affected, other nodes will compensate for the spread, and non-central nodes are more likely to receive information.

    \item Algorithm optimization.

    In addition to sample collection, we can optimize algorithmically. There are many other ways to try now. For binary time series, we can try other ways by selecting only the $t_0$ and $t_1$ times of the time series matrix. Repeat the experiment several times, compare the infection status of multiple groups at $t_0$ and $t_1$, and infer the connection of nodes. If we want to get out of the category of binary time series, we can also choose neural networks. For example, the hyper-graph neural network (HGNN) method \cite{Bai2021, Huang2021, Young2021}. Although graph neural networks (GNNs) have achieved success in graph representation learning, it remains a challenging task how to directly apply powerful GNN variants to hypergraphs. This is also a direction for research on similar issues in the future.
\end{enumerate}

The above three solutions are the methods that can optimize the results when dealing with real problems when we know that the algorithm in this paper is most suitable for small-world networks and the shortcomings of the algorithm itself. The purpose of the first two solutions is to find a way to make our sample network closer to the small-world network, whose main purpose is to reduce the degree variance. The third solution is to find a breakthrough point in the algorithm when it is inconvenient to rectify the sample. When the propagation time is limited and the disturbance is not easy to control, these three methods are feasible ideas for optimizing our experiments.

In section 4, we can conclude that due to the flattening of the small world network, the information dissemination efficiency is the highest, and the network reconstruction method in this paper is the most effective. If we want to experiment smoothly, we can make the network model closer to the small-world network. If the structure cannot be changed, we can only try new methods.

\section{Conclusion}
In this paper, the maximum likelihood estimation is introduced into complex networks, mainly based on SIS-type propagation dynamics with binary state dynamic variables on the complex network composed of simplex, namely simplex complex. To reconstruct the complex network, we infer the existence of connected edges and simplex according to the change of node infection state with a long period of propagation. In the whole process of machine learning, we use vectorization operations to infer the adjacency matrix of the network.

We first simulate the propagation process according to the SIS propagation model to generate matrices $\Psi_1=(\psi_i^{t_m})_{T\times n}$ and $\Psi_2=(\psi_{jk}^{t_m})_{T\times n_1}$ which represent the time series of infection status. Next, we use the combination of the probabilities $P^i_j$, $P^i_{jk}$ and the EM algorithm to infer the probabilities $P_{j\to i}$, $P_{jk\to i}$ of propagation on 1-simplex and 2-simplex. We can put all these results into several matrices. In order to confirm the existence of simplex, we need to redefine a threshold for judging the existence of simplex, in the meantime using the $F1$ score to quantify the effectiveness of complex network reconstruction. After experimenting on four kinds of complex networks, we judge that the reduction degree of 1-simplex complex can satisfy $F1>0.8$ when $T\ge8000$ according to the experimental data, while more iterations are required under the two-step iterative method for the 2-simplicial complex. For simplicial complexes, $F1$ increases as the number of iterations $T$ increases, and it decreases as the perturbation $\rho$ increases.

For different networks, the reconstruction effectiveness is different when using the method simulated in this paper. Through comparison, we find that the larger the degree variance $\sigma^2$ of the network type, the lower the effectiveness of the network reconstruction when the number of nodes is the same and the average degree is in the same order of magnitude, under the same number of iterations. For networks with the same number of nodes and the average degree of the same order of magnitude, a large difference in degree variance means a large difference in network structure. In order to test, we also introduce the relationship between reconfiguration effectiveness and mutual information of network node degrees. As well as the betweenness centrality of the seven networks and their corresponding reconstruction effectiveness.

In this paper, we discover that the small-world network is the most suitable type of network for using the experimental method, and this kind of network meets real networks such as human-to-human social networks. When the time is limited and the external environment cannot be changed, there are three methods to improve the effectiveness of the experiment. The first and second methods are to decentralize and extend the scale of experimental network samples so that the network is closer to the small-world network. The third method is to use new methods such as HGNN to maintain the validity of the experiment when the network structure cannot be changed.

The contribution of this paper has four aspects, the vectorized expression of network information dissemination variables,  inferring the relationship between reconstruction effectiveness and network type, using the difference of degree variance as an index to evaluate the difference of network structure, and the optimization methods of making the network closer to small-world network are presented.

That look for a more optimized method and a quantifiable unified standard to judge network types is the direction we can further study. The directions for further research are as follows:
\begin{enumerate}[(1)]
    \item Try network reconstruction with HGNN and compare it with EM algorithm network reconstruction.
    \item Try to introduce the E-Bayesian estimation into the estimation of the network adjacency matrix, and the reconstruction under the EM algorithm can be regarded as a comparison.
    \item Replace the time series matrix with experimental data that only propagates from time $t_0$ to time $t_1$ while repeating it many times to test the effectiveness of reconstruction.
    \item Looking for quantitative indicators to judge the difference between the network structure and the small-world network in addition to the degree variance.
    \item Combining the whole theme with casual emergence and coarse graining can be a new idea.
\end{enumerate}

\section*{Acknowledgement}
Thank the reviewers for their valuable comments. We have followed the advice and improved some of the arguments, so that readers can understand the specific derivation process more clearly. In response to the reviewers' comments and suggestions, we have added appendixes to the paper, which can make the description clear and provide a trail for readers who want to study it in depth.

\appendix
\renewcommand{\thesection}{Appendix \Alph{section}}
\renewcommand\theequation{\Alph{section}.\arabic{equation}} 
\section[\appendixname~\thesection]{Simplex and dynamics}\label{simple}

\begin{figure}[htbp]
	\begin{minipage}[t]{\linewidth}
		\centering
		\includegraphics[width=0.8\linewidth]{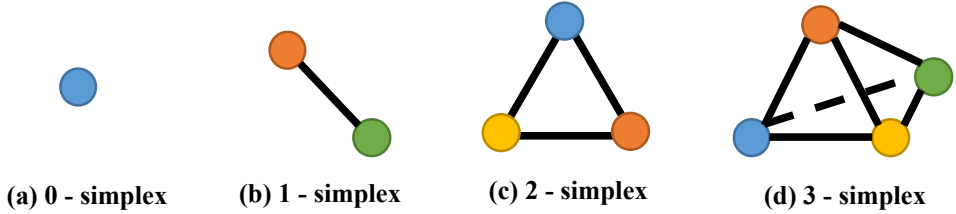}
	\end{minipage}
	\caption{A 0-simplex in (a) represents a single node, a 1-simplex in (b) represents a node pair and its connected edge, a 2-simplex in (c) represents a fully connected triangle structure composed of three nodes, and a 3-simplex in (d) represents a fully connected tetrahedron structure composed of four nodes.}
	\label{simplex}
\end{figure}

\begin{figure}[htbp]
	\begin{minipage}[t]{\linewidth}
		\centering
		\includegraphics[width=0.8\linewidth]{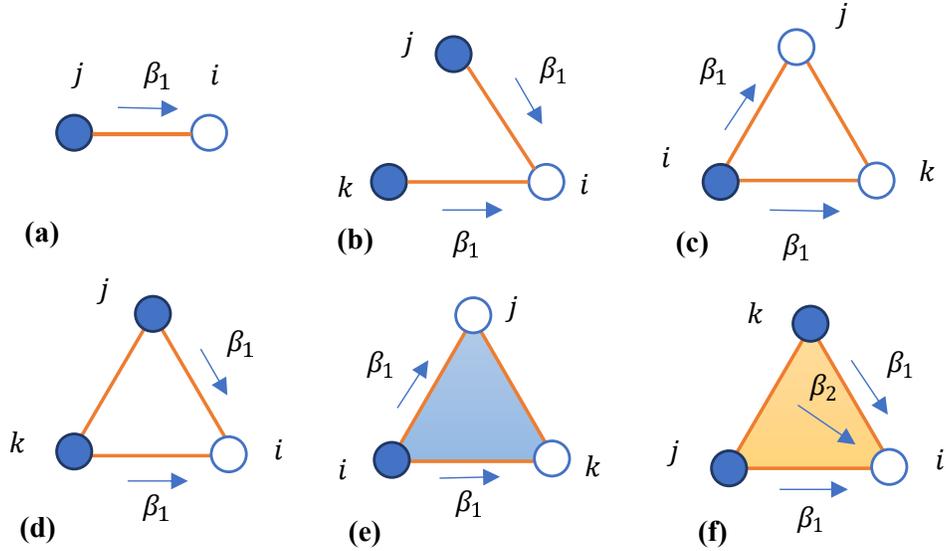}
	\end{minipage}
	\caption{(a)-(f) represent the propagation rule of infectious state under 6 conditions, blue nodes represent infectious state nodes, and white nodes represent susceptible state nodes. $\beta_1$ and $\beta_2$ indicate the probability of the infectious state spreading along the arrow direction. As an ordinary connection (a), there is only the effect of two-body propagation. There are three nodes in (b), but there is no connecting edge between points $j$ and $k$, so it does not form a $2$-simplex and there is only node-to-node propagation. Although there are conditions to form a $2$-simplex in (c) and (d), there is no simplex property between the three nodes, so even if $j$ and $k$ in (d) are infected, three-body propagation won't happen. There are simplexes in the three nodes in (e) and (f), while in (f), there happens to be that $j$ and $k$ are both in the infectious state. So there is a $1$-simplex $jk$, which can spread infectious state to $i$ through the simplex, and the propagation probability is $\beta_2$.}
	\label{inter}
\end{figure}

The $K$-simplex describes the simultaneous interaction between $k+1$ nodes. Any higher-order simplex can be disassembled into several lower-order simplexes. According to the definition, the $k$-order simplex can be disassembled into $(k-1)$-simplex whose quantity is $k+1$. The interaction on a $k$-simplex can be defined as the effect of the $(k-1)$-simplex on a single node. The network composed of the simplex is called the simplex complex. An article on simplex for social models \cite{Iacopini2019} details the definition of simplex and its relationship to hypergraphs. The detailed definition of simplex and the dynamic propagation mechanism on it \cite{Iacopini2019, Battiston2020} can be viewed in the literature on propagation dynamics on simplex.  We summarized the contents of them. From Fig.\ref{simplex} we can see that a 0-simplex represents a single node, a 1-simplex represents a node pair and its connected edge, a 2-simplex represents a fully connected triangle structure composed of three nodes, and a 3-simplex represents a fully connected tetrahedron structure composed of four nodes. From Fig.\ref{inter} we can learn the propagation rule of infectious state under 6 conditions. Only when a simplex exists and at least one lower order simplex contained in it is infected, the infectious state transmission on the simplex will occur.

\section[\appendixname~\thesection]{MLE derivation process}\label{ML}
For experimental data, when $\psi_i^{t_m}=1,\psi_j^t=1,\psi_i^t=0$, we stipulate that $j$ may transmit infectious state to $i$. The propagation possibility of a $1$-simplex to a node is similar.  When $\psi_i^{t+1}=1,\psi_k^t=1,\psi_j^t=1,\psi_i^t=0$, we stipulate that $jk$ may transmit infectious state to $i$. After conditions are met, we also need to judge whether $\psi_i^{t+1}=1$ is caused by $\psi_j^t=1$ or $\psi_{jk}^t=1$. We express this situation as $j\to i$ or $jk\to i$. From this, we can conclude whether there is a simplex between $i$ and $j$ or $jk$. 

We get the probability of $\psi_i^{t_m}=1,\psi_j^t=1,\psi_i^t=0$, $P(\psi_i^{t_m}=1|\psi_j^t=1,\psi_i^t=0)$, from the experimental data. The probability is presented by the ratio of the time nodes that meet the conditions to the total number of iterations. We abbreviate it as
\begin{eqnarray}
	P^i_j=P(\psi_i^{t_m}=1|\psi_j^t=1,\psi_i^t=0),
\end{eqnarray}
in the same way, $P(\psi_i^{t+1}=1|\psi_k^t=1,\psi_j^t=1,\psi_i^t=0)$ is presented as
\begin{eqnarray}
	P^i_{jk}=P(\psi_i^{t+1}=1|\psi_k^t=1,\psi_j^t=1,\psi_i^t=0).
\end{eqnarray}

In order to judge the existence of the connected edge or simplex, we take two groups of variables, $P_{j\to i}$ and $P_{jk\to i}$ as the main training variables. We define two sets of variables as
\begin{eqnarray}
   P_{j\to i}=P(j\to i|\psi_i^{t+1}=1,\psi_j^t=1,\psi_i^t=0)
\end{eqnarray}
and
\begin{eqnarray}
	P_{jk\to i}=P(j\to i|\psi_i^{t+1}=1,\psi_j^t=1, \psi_k^t=1,\psi_i^t=0).
\end{eqnarray}
In which $i,j,k=1,2,...,n, i\neq j\neq k$.

According to the concept of conditional probability, we can get the following two formulas as
\begin{eqnarray}
	P_{\psi_i|{j}}=P(j\to i,\psi_i^{t+1}=1|\psi_j^t=1,\psi_i^t=0)=P^i_j P_{j\to i}
\end{eqnarray}
and
\begin{eqnarray}
	P_{\psi_i|{jk}}=P(jk\to i,\psi_i^{t+1}=1|\psi_k^t=1, \psi_j^t=1,\psi_i^t=0)=P^i_{jk} P_{jk\to i}.
\end{eqnarray}

Therefore, we can calculate the infection state of $i$ at time $t$ according to the mathematical expectation formula as
\begin{eqnarray}\label{exppsi}
	\begin{aligned}
		E(\psi_i^{t+1})&=1\times P(\psi_i^{t+1}=1)+0\times P(\psi_i^{t+1}=0)+\varepsilon_i=P(\psi_i^{t+1}=1)+\varepsilon_i\\&
		=\displaystyle\sum_{j,(j\ne i)}P_{\psi_i|{j}}\psi_j^{t}+\displaystyle\sum_{j,k,(j,k\ne i)}P_{\psi_i|{jk}}\psi_{j}^{t}\psi_{k}^{t}+\varepsilon_i\\&
		=\displaystyle\sum_{j,(j\ne i)}P^i_j P_{j\to i}\psi_j^{t}+\displaystyle\sum_{j,k,(j,k\ne i)}P^i_{jk} P_{jk\to i}\psi_{k}^{t}\psi_{j}^{t}+\varepsilon_i.
	\end{aligned}
\end{eqnarray}
Making data real can make machine learning more effective. In order to make the machine learning fit better, we introduce the noise variable $\varepsilon_i$. $i,j,k=1,2,...,n, i\neq j\neq k$.

By introducing logarithm and Jason inequality, we can obtain the inequality of $i$ state expectation at time $t+1$. In the calculation, we introduce three intermediate quantities, $\rho_{i,j}^t, \rho_{i,jk}^t$ and $\rho_{\varepsilon_i}^t$. Three variables are used in the iteration.
\begin{eqnarray}\label{logexppsi}
	\begin{aligned}
		\ln E(\psi_i^{t+1})=&\ln\left(\displaystyle\sum_{j,(j\ne i)}P^i_j P_{j\to i}\psi_j^{t}+\displaystyle\sum_{j,k,(j,k\ne i)}P^i_{jk} P_{jk\to i}\psi_{k}^{t}\psi_{j}^{t}+\varepsilon_i\right)\\=&\ln\left(\displaystyle\sum_{j,(j\ne i)}\rho_{i,j}^t\frac{P^i_j P_{j\to i}\psi_j^{t}}{\rho_{i,j}^t}+\displaystyle\sum_{j,k,(j,k\ne i)}\rho_{i,jk}^t\frac{P^i_{jk} P_{jk\to i}\psi_j^{t}\psi_k^{t}}{\rho_{i,jk}^t}+\rho_{\varepsilon_i}^t\frac{\varepsilon_i}{\rho_{\varepsilon_i}^t}\right)\\\geq &\displaystyle\sum_{j,(j\ne i)}\rho_{i,j}^t\ln \frac{P^i_j P_{j\to i}\psi_j^{t}}{\rho_{i,j}^t}+\displaystyle\sum_{j,k,(j,k\ne i)}\rho_{i,jk}^t \ln \frac{P^i_{jk} P_{jk\to i}\psi_j^{t}\psi_k^{t}}{\rho_{i,jk}^t}+\rho_{\varepsilon_i}^t\ln \frac{\varepsilon_i}{\rho_{\varepsilon_i}^t}\\= &\displaystyle\sum_{j,(j\ne i)}\rho_{i,j}^t\ln P^i_j P_{j\to i}\psi_j^{t}+\displaystyle\sum_{j,k,(j,k\ne i)}\rho_{i,jk}^t \ln P^i_{jk} P_{jk\to i}\psi_j^{t}\psi_k^{t}+\rho_{\varepsilon_i}^t\ln \varepsilon_i\\&-\displaystyle\sum_{j,(j\ne i)}\rho_{i,j}^t\ln \rho_{i,j}^t+\displaystyle\sum_{j,k,(j,k\ne i)}\rho_{i,jk}^t \ln \rho_{i,jk}^t+\rho_{\varepsilon_i}^t\ln \rho_{\varepsilon_i}^t.
	\end{aligned}.
\end{eqnarray}
When the equal sign is established, the three intermediate values meet
\begin{eqnarray}\label{rhoij}
	\rho_{i,j}^t=\frac{P^i_j P_{j\to i}\psi_j^{t}}{\displaystyle\sum_{j,(j\ne i)}P^i_j P_{j\to i}\psi_j^{t}+\displaystyle\sum_{j,k,(j,k\ne i)}P^i_{jk} P_{jk\to i}\psi_{k}^{t}\psi_{j}^{t}+\varepsilon_i},
\end{eqnarray}
\begin{eqnarray}\label{rhoijk}
	\rho_{i,jk}^t=\frac{P^i_{jk} P_{jk\to i}\psi_{k}^{t}\psi_{j}^{t}}{\displaystyle\sum_{j,(j\ne i)}P^i_j P_{j\to i}\psi_j^{t}+\displaystyle\sum_{j,k,(j,k\ne i)}P^i_{jk} P_{jk\to i}\psi_{k}^{t}\psi_{j}^{t}+\varepsilon_i},
\end{eqnarray}
and
\begin{eqnarray}\label{rhoeps}
	\rho_{\varepsilon_i}^t=\frac{\varepsilon_i}{\displaystyle\sum_{j,(j\ne i)}P^i_j P_{j\to i}\psi_j^{t}+\displaystyle\sum_{j,k,(j,k\ne i)}P^i_{jk} P_{jk\to i}\psi_{k}^{t}\psi_{j}^{t}+\varepsilon_i}.
\end{eqnarray}

Next, we use maximum likelihood estimation and expectation maximization for operation. The probability of a given number of events occurring in a fixed time interval is represented by Poisson distribution, so we use Poisson distribution to capture the randomness of node $i$'s infection time \cite{Karrer2011,Newman2016,DeBacco2017,Wang2022}. Poisson distribution can make the analysis and calculation of EM algorithm more convenient. In this paper, the expectation about the infection state of node $i$ is expressed as
\begin{eqnarray}
	P(\Psi_i^{t+1}=\psi_i^{t+1})=\frac{e^{-E(\psi_i^{t+1})}[E(\psi_i^{t+1})]^{\psi_i^{t+1}}}{\psi_i^{t+1} !}.
\end{eqnarray}
Combining the probabilities of the states of all effective nodes, we can obtain the maximum likelihood estimate of $P_{j\to i},P_{jk\to i}$ and $\varepsilon_i$. We can get the likelihood function through probability, which is expressed as
\begin{eqnarray}\label{likelihood}
	\begin{aligned}
		L(P_{j\to i}, P_{jk\to i}, \varepsilon_i)&=\displaystyle\prod_{m(\psi_i^t=0)}P(\Psi_i^{t+1}=\psi_i^{t+1}) \propto e^{-E(\psi_i^{t+1})}[E(\psi_i^{t+1})]^{\Psi_i^{t+1}}=\widetilde{L}(P_{j\to i}, P_{jk\to i}, \varepsilon_i)
	\end{aligned}.
\end{eqnarray}
In \ref{likelihood}, $m(\psi_i^t=0)$ means to judge whether $\psi_i^t=0$. If it is unqualified, $P(\Psi_i^{t+1}=\psi_i^{t+1})$ will not be included in the function. Take the logarithm on both sides of the equal sign,
\begin{eqnarray}\label{lnl}
	\begin{aligned}
		\ln\widetilde{L}(P_{j\to i}, P_{jk\to i}, \varepsilon_i)&=\displaystyle\sum_{m(\psi_i^t=0)}\psi_i^{t+1}\ln E(\psi_i^{t+1})-E(\psi_i^{t+1})
	\end{aligned}
\end{eqnarray}
Then we make the partial derivative of the log-likelihood function with respect to the three groups of variables equal to 0,
\begin{eqnarray}
\frac{\partial\ln\widetilde{L}}{\partial P_{j\to i}}=0,
\end{eqnarray}
\begin{eqnarray}
	\frac{\partial\ln\widetilde{L}}{\partial P_{jk\to i}}=0,
\end{eqnarray}
and
\begin{eqnarray}
	\frac{\partial\ln\widetilde{L}}{\partial \varepsilon_i}=0.
\end{eqnarray}
Combined with the formula (\ref{exppsi}), (\ref{logexppsi}) and (\ref{lnl}), we can get
\begin{eqnarray}\label{pjtoi}
	P_{j\to i}=\frac{\displaystyle\sum_{m(\psi_i^t=0)}\rho_{i,j}^t \psi_i^{t+1}}{\displaystyle\sum_{m(\psi_i^t=0)}P^i_j \psi_j^{t}},
\end{eqnarray}
\begin{eqnarray}\label{pjktoi}
	P_{jk\to i}=\frac{\displaystyle\sum_{m(\psi_i^t=0)}\rho_{i,jk}^t \psi_i^{t+1}}{\displaystyle\sum_{m(\psi_i^t=0)}P^i_{jk} \psi_{jk}^{t}},
\end{eqnarray}
and
\begin{eqnarray}\label{epsi}
	\varepsilon_i=\frac{\displaystyle\sum_{m(\psi_i^t=0)}\rho_{\varepsilon_i}^t \psi_i^{t+1}}{\displaystyle\sum_{m(\psi_i^t=0)}1}.
\end{eqnarray}
In this way, we can find the iteration variables from (\ref{rhoij})-(\ref{rhoeps}) and (\ref{pjtoi})-(\ref{epsi}). Although this expression is reasonable, it is still inappropriate to use machine learning language. Because in Python, we need to use a large number of \emph{for} loops to repeatedly calculate these variables. The operation efficiency will be greatly reduced. Therefore, we will vectorize six groups of variables, and the vector and matrix can turn the \emph{for} loop into parallel operation.

For the two-step method, we only need to omit the multi-body effect in the first step. The expectation is
\begin{eqnarray}\label{exppsi0}
	\begin{aligned}
		E_0(\psi_i^{t+1})&=1\times P(\psi_i^{t+1}=1)+0\times P(\psi_i^{t+1}=0)+\varepsilon_i=P(\psi_i^{t+1}=1)+\varepsilon_i\\&
		=\displaystyle\sum_{j,(j\ne i)}P_{\psi_i|{j}}\psi_j^{t}+\varepsilon_i
		=\displaystyle\sum_{j,(j\ne i)}P^i_j P_{j\to i}\psi_j^{t}+\varepsilon_i.
	\end{aligned}
\end{eqnarray}

The intermediate variables in the first step are
\begin{eqnarray}\label{rho0ij}
	\rho_{0i,j}^t=\frac{P^i_j P^0_{j\to i}\psi_j^{t}}{\displaystyle\sum_{j,(j\ne i)}P^i_j P^0_{j\to i}\psi_j^{t}+\varepsilon_i},
\end{eqnarray}
and
\begin{eqnarray}\label{rho0eps}
	\rho_{\varepsilon_{0i}}^t=\frac{\varepsilon_i}{\displaystyle\sum_{j,(j\ne i)}P^i_j P^0_{j\to i}\psi_j^{t}+\varepsilon_i}.
\end{eqnarray}

The variables of learning are
\begin{eqnarray}\label{p0jtoi}
	P^0_{j\to i}=\frac{\displaystyle\sum_{m(\psi_i^t=0)}\rho_{i,j}^t \psi_i^{t+1}}{\displaystyle\sum_{m(\psi_i^t=0)}P^i_j \psi_j^{t}},
\end{eqnarray}
and
\begin{eqnarray}\label{eps0i}
	\varepsilon_{0i}=\frac{\displaystyle\sum_{m(\psi_i^t=0)}\rho_{\varepsilon_i}^t \psi_i^{t+1}}{\displaystyle\sum_{m(\psi_i^t=0)}1}.
\end{eqnarray}
\section[\appendixname~\thesection]{Vectorized representations and single-element representation transformations}\label{vector}
This appendix shows the main matrices in the text. $(jk)_l$ represents the $l$-th $1$-simplex in the network, $l=1,\ldots,n_1$. For each simplex, we rank the nodes with smaller numbers first to get $(j,k)$. We sort $j$ from small to large, and then we sort $k$ from small to large after each group of $j$. For example, $(1,4)$, $(2,5)$, $(1,5)$ and $(3,4)$, we sort them as $(1,4)$, $(1,5)$, $(2,5)$ and $(3,4)$. 

We use matrix $\Psi_1=(\psi_i^{t_m})_{T\times n}$ to represent the time series of infection status of the whole network or simplex complex. On the basis of matrix $\Psi_1$, we can generate the matrix of infection state of 1-simplex, which is represented by time series matrix $\Psi_2=(\psi_{jk}^{t_m})_{T\times n_1}$. If both nodes $j$ and $k$ are in the infected state, we specify $\psi_{jk}^{t_m}=1$. The specific structure of the two matrices is
\begin{eqnarray}
	\Psi_1=
	\left(\begin{array}{cccc}
		\psi^1_1 & \psi^1_2 &\ldots & \psi^1_n\\
		\psi^2_1 & \psi^2_2 & \ldots & \psi^2_n\\
		\ldots & \ldots & \ldots & \ldots\\ 
		\psi^T_1 & \psi^T_2 & \ldots & \psi^T_n\\ 
	\end{array}\right)_{T\times n},
	\Psi_2=
\left(\begin{array}{cccc}
	\psi^1_{{(jk)}_1} & \psi^1_{{(jk)}_2} &\ldots & \psi^1_{{(jk)}_{n_1}}\\
	\psi^2_{{(jk)}_1} & \psi^2_{{(jk)}_2} & \ldots & \psi^2_{{(jk)}_{n_1}}\\
	\ldots & \ldots & \ldots & \ldots\\ 
	\psi^T_{{(jk)}_1} & \psi^T_{{(jk)}_2} & \ldots & \psi^T_{{(jk)}_{n_1}}\\ 
\end{array}\right)_{T\times n_1}.
\end{eqnarray}

In order to meet the required expression, we introduce matrices $Pt_1=(P_j^i)_{n\times n}$ and $Pt_2=(P_{jk}^i)_{n\times n_1}$. The specific results of the two matrices are shown as
\begin{eqnarray}
Pt_1=
\left(\begin{array}{cccc}
	P^1_1 & P^1_2 &\ldots & P^1_n\\
	P^2_1 & P^2_2 & \ldots & P^2_n\\
	\ldots & \ldots & \ldots & \ldots\\ 
	P^n_1 & P^2_2 & \ldots & P^n_n\\ 
\end{array}\right)_{n\times n},
 Pt_2=
\left(\begin{array}{cccc}
	P^1_{{(jk)}_1} & P^1_{{(jk)}_2} &\ldots & P^1_{{(jk)}_{n_1}}\\
	P^2_{{(jk)}_1} & P^2_{{(jk)}_2} & \ldots & P^2_{{(jk)}_{n_1}}\\
	\ldots & \ldots & \ldots & \ldots\\ 
	P^n_{{(jk)}_1} & P^2_{{(jk)}_2} & \ldots & P^n_{{(jk)}_{n_1}}\\ 
\end{array}\right)_{n\times n_1}.
\end{eqnarray}

Write $P_{j\to i}$ and $P_{jk\to i}$ into the matrix, expressed as
\begin{eqnarray}
	P_1=
	\left(\begin{array}{cccc}
		P_{1\to1} & P_{1\to2} &\ldots &P_{n\to n}\\
		P_{2\to1} & P_{1\to2} & \ldots & P_{n\to n}\\
		\ldots & \ldots & \ldots & \ldots\\ 
		P_{n\to1} & P_{n\to1} & \ldots & P_{n\to n}\\ 
	\end{array}\right)_{n\times n},
	 P_2=
	\left(\begin{array}{cccc}
		P_{{(jk)}_1\to 1} & P_{{(jk)}_1\to 1} &\ldots & P^1_{{(jk)}_{n_1}}\\
		P_{{(jk)}_2\to 1} & P_{{(jk)}_2\to 2} & \ldots & P^2_{{(jk)}_{n_1}}\\
		\ldots & \ldots & \ldots & \ldots\\ 
		P_{{(jk)}_{n_1}\to 1} & P_{{(jk)}_{n_1}\to 2} & \ldots & P_{{(jk)}_{n_1}\to n}\\ 
	\end{array}\right)_{n_1\times n}.
\end{eqnarray}

To make the data more realistic, we introduce noise
\begin{eqnarray}
	Ep=(\varepsilon_1,\varepsilon_2,\ldots,\varepsilon_n)^T
\end{eqnarray}

\begin{figure}[htbp]
	\begin{minipage}[t]{\linewidth}
		\centering
		\includegraphics[width=0.8\linewidth]{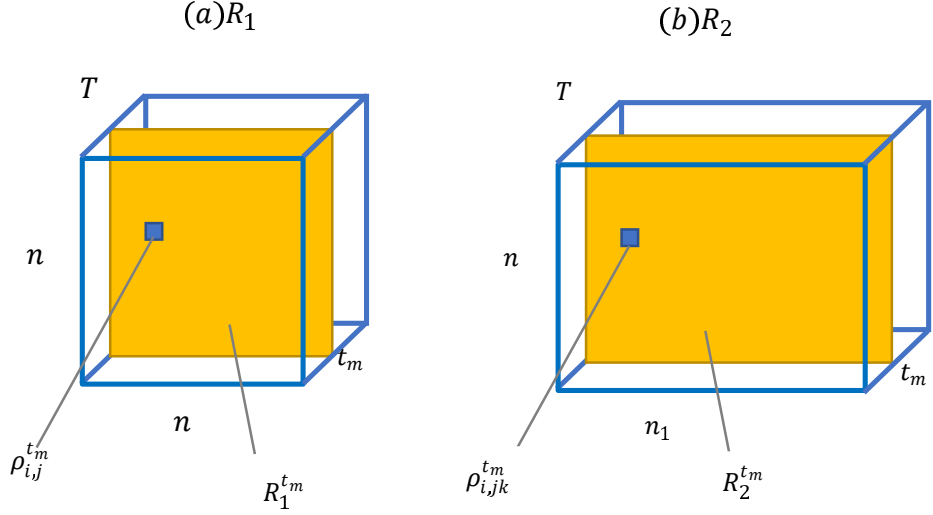}
	\end{minipage}
	\caption{Schematic diagrams of three-dimensional matrices (a)$R_1$ and (b)$R_2$. $R_1^{t_m}$ and $R_2^{t_m}$ like two sections of two boxes at point $t_m$. 	$\rho_{i,j}^{t_m}$ and 	$\rho_{i,jk}^{t_m}$ are the elements on the cross sections.}
	\label{rho}
\end{figure}

Since the intermediate variable has three dimensions, we take the matrix at time $t_m$ for display as
\begin{eqnarray}
	R_1^{t_m}=
	\left(\begin{array}{cccc}
		\rho_{1,1}^{t_m} &\rho_{1,2}^{t_m} &\ldots & \rho_{1,n}^{t_m}\\
		\rho_{2,1}^{t_m} & \rho_{2,2}^{t_m} & \ldots & \rho_{2,n}^{t_m}\\
		\ldots & \ldots & \ldots & \ldots\\ 
		\rho_{n,1}^{t_m} & \rho_{n,2}^{t_m} & \ldots & \rho_{n,n}^{t_m}\\ 
	\end{array}\right)_{n\times n},
	R_2^{t_m}=
	\left(\begin{array}{cccc}
		\rho_{1,{(jk)}_1}^{t_m} & \rho_{1,{(jk)}_2}^{t_m}  &\ldots & \rho_{1,{(jk)}_{n_1}}^{t_m} \\
	\rho_{2,{(jk)}_1}^{t_m}  & \rho_{2,{(jk)}_2}^{t_m}  & \ldots & \rho_{2,{(jk)}_{n_1}}^{t_m} \\
		\ldots & \ldots & \ldots & \ldots\\ 
		\rho_{n,{(jk)}_1}^{t_m}  & \rho_{n,{(jk)}_2}^{t_m}  & \ldots & \rho_{n,{(jk)}_{n_1}}^{t_m} \\ 
	\end{array}\right)_{n\times n_1}.
\end{eqnarray}
In this way, the three matrices $R_{\varepsilon}, R_1$ and $R_2$ can be expressed as
\begin{eqnarray}
	R_{\varepsilon}=	\left(\begin{array}{cccc}
		\rho_{\varepsilon_1}^{t_1} &\rho_{\varepsilon_2}^{t_1} &\ldots & \rho_{\varepsilon_n}^{t_1}\\
		\rho_{\varepsilon_1}^{t_2} & \rho_{\varepsilon_2}^{t_2} & \ldots & \rho_{\varepsilon_n}^{t_2}\\
		\ldots & \ldots & \ldots & \ldots\\ 
		\rho_{\varepsilon_1}^{t_T} & \rho_{\varepsilon_2}^{t_T} & \ldots & \rho_{\varepsilon_n}^{t_T}\\ 
	\end{array}\right)_{n\times T},
\end{eqnarray}

\begin{eqnarray}
	R_1=(R_1^{t_1},	R_1^{t_2},\ldots,R_1^{t_T})_{n\times n\times T}
\end{eqnarray}
and
\begin{eqnarray}
	R_2=(R_2^{t_1},	R_2^{t_2},\ldots,R_2^{t_T})_{n\times n_1\times T}.
\end{eqnarray}
$R_1$ and $R_2$ are three-dimensional matrices. Schematic diagrams of three-dimensional matrices $R_1$ and $R_2$ are shown in Fig.\ref{rho}.

Combined with (\ref{rhoij})-(\ref{rhoeps}) and (\ref{pjtoi})-(\ref{epsi}), we can get the matrix iteration formula in the text through the above matrix.

For two-step iterative method,
\begin{eqnarray}
	P_0=
	\left(\begin{array}{cccc}
		P^0_{1\to1} & P^0_{1\to2} &\ldots &P^0_{n\to n}\\
		P^0_{2\to1} & P^0_{1\to2} & \ldots & P^0_{n\to n}\\
		\ldots & \ldots & \ldots & \ldots\\ 
		P^0_{n\to1} & P^0_{n\to1} & \ldots & P^0_{n\to n}\\ 
	\end{array}\right)_{n\times n},
\end{eqnarray}
\begin{eqnarray}
	Ep_0=(\varepsilon_{01},\varepsilon_{02},\ldots,\varepsilon_{0n})^T.
\end{eqnarray}
At time $t_m$,
\begin{eqnarray}
	R_0^{t_m}=
	\left(\begin{array}{cccc}
		\rho_{01,1}^{t_m} &\rho_{01,2}^{t_m} &\ldots & \rho_{01,n}^{t_m}\\
		\rho_{0,1}^{t_m} & \rho_{02,2}^{t_m} & \ldots & \rho_{02,n}^{t_m}\\
		\ldots & \ldots & \ldots & \ldots\\ 
		\rho_{0n,1}^{t_m} & \rho_{0n,2}^{t_m} & \ldots & \rho_{0n,n}^{t_m}\\ 
	\end{array}\right)_{n\times n},
\end{eqnarray}
In the same way,
\begin{eqnarray}
	R_{\varepsilon0}=	\left(\begin{array}{cccc}
		\rho_{\varepsilon_01}^{t_1} &\rho_{\varepsilon_02}^{t_1} &\ldots & \rho_{\varepsilon_0n}^{t_1}\\
		\rho_{\varepsilon_01}^{t_2} & \rho_{\varepsilon_02}^{t_2} & \ldots & \rho_{\varepsilon_0n}^{t_2}\\
		\ldots & \ldots & \ldots & \ldots\\ 
		\rho_{\varepsilon_01}^{t_T} & \rho_{\varepsilon_02}^{t_T} & \ldots & \rho_{\varepsilon_0n}^{t_T}\\ 
	\end{array}\right)_{n\times T},
\end{eqnarray}

\begin{eqnarray}
	R_0=(R_0^{t_1},	R_0^{t_2},\ldots,R_0^{t_T})_{n\times n_1\times T}.
\end{eqnarray}
In this way, we can get the matrix of the two-step iterative method.

\bibliographystyle{ieeetr}
\bibliography{jiaoda11}

\end{document}